\documentclass{svjour3}
\usepackage{lineno,hyperref}

\usepackage{graphicx}
\usepackage{amssymb,amsmath}
\usepackage{verbatim}
\usepackage[caption=false, justification=centering]{subfig}
\usepackage[numbers]{natbib}
\usepackage[nolist]{acronym}

\AtBeginDocument{%
  \setlength{\oddsidemargin}{\dimexpr(\paperwidth-\textwidth)/2-1in}%
  \setlength{\evensidemargin}{\oddsidemargin}%
  \setlength{\topmargin}{%
    \dimexpr(\paperheight-\textheight)/2-\headheight-\headsep-1in}%
}

\begin{document}
\title{Sequence-based Detection of Sleeping Cell Failures in Mobile Networks}



\author{Fedor Chernogorov         \and
        Sergey Chernov \and
        Kimmo Brigatti \and
        Tapani Ristaniemi 
}


\institute{F. Chernogorov \at
              Magister Solution Ltd., Sepankatu 14 C, FIN-40720, 		Jyvaskyla, Finland \\
              University of Jyvaskyla, Department of Mathematical Information Technology, P.O. Box 35, FI-40014 University of Jyvaskyla, Finland
              \email{fedor.chernogorov[at]magister.fi, fedor.chernogorov[at]jyu.fi}
\and
           S. Chernov, K. Brigatti and Tapani Ristaniemi \at
			University of Jyvaskyla, Department of Mathematical Information Technology, P.O. Box 35, FI-40014 University of Jyvaskyla, Finland
              \email{sergey.a.chernov[at]jyu.fi, kimmobrigatti[at]gmail.com, tapani.e.ristaniemi[at]jyu.fi}
}
\date{Received: date / Accepted: date}

\begin{acronym}[eutrannnnnn]

\acro{2g}[2G]{$2^{nd}$ Generation}
\acro{3g}[3G]{$3^{rd}$ Generation}
\acro{3gpp}[3GPP]{$3^{rd}$ Generation Partnership Programme}
\acro{4g}[4G]{$4^{th}$ Generation}
\acro{5g}[5G]{$5^{th}$ Generation}
\acro{aa}[AA]{Anomaly Analysis}
\acro{agnes}[AGNES]{AGglomerative NESting}
\acro{agnss}[A-GNSS]{Assisted-Global Navigation Satellite System}
\acro{anr}[ANR]{Automatic Neighbor Relations}
\acro{auc}[AUC]{Area under Curve}
\acro{bcr}[BCR]{Blocked Call Rate}
\acro{ber}[BER]{Bit Error Rate}
\acro{birch}[BIRCH]{Balanced Iterative Reducing and Clustering Using Hierarchies}
\acro{bler}[BLER]{Block Error Rate}
\acro{bs}[BS]{Base Station}
\acro{cblof}[CBLOF]{Cluster-Based Local Outlier Factor}
\acro{cbr}[CBR]{Case-Based Reasoning}
\acro{ccsr}[CCSR]{Call Completion Success Rate}
\acro{clique}[CLIQUE]{CLustering In QUEst}
\acro{cm}[CM]{Configuration Management}
\acro{coc}[COC]{Cell Outage Compensation}
\acro{cod}[COD]{Cell Outage Detection}
\acro{commune}[COMMUNE]{COgnitive network ManageMent under UNcErtainty}
\acro{cpich}[CPICH]{Common Pilot Channel}
\acro{cqi}[CQI]{Channel Quality Indicator}
\acro{crnti}[C-RNTI]{Cell Radio Network Temporary Identifier}
\acro{csi}[CSI]{Channel State Indicator}
\acro{cssr}[CSSR]{Call Setup Success Rate}
\acro{dbscan}[DBSCAN]{Density-based spatial clustering of applications with noise}
\acro{dcr}[DCR]{Drop Call Ratio}
\acro{diana}[DIANA]{DIvisive ANAlysis}
\acro{dl}[DL]{Downlink}
\acro{dm}[DM]{Diffusion Maps}
\acro{drx}[DRX]{Discontinuous Reception}
\acro{enb}[eNB]{E-UTRAN NodeB}
\acro{eps}[EPS]{Evolved Packet System}
\acro{eutran}[E-UTRAN]{Evolved Universal Terrestrial Radio Access Network}
\acro{fdd}[FDD]{Frequency Division Duplexing}
\acro{fer}[FER]{Frame Error Rate}
\acro{fm}[FM]{Fault Management}
\acro{fpr}[FPR]{False Positive Rate}
\acro{frf}[FRF]{Frequency Reuse Factor}
\acro{gsm}[GSM]{Global System for Mobile Communications}
\acro{harq}[HARQ]{Hybrid Adaptive Repeat and reQuest}
\acro{hdp}[HDP]{Hierarchical Dirichlet Process}
\acro{hlr}[HLR]{Home Location Register}
\acro{ho}[HO]{Handover}
\acro{hof}[HOF]{Handover Failure}
\acro{hspa}[HSPA]{High Speed Packet Access}
\acro{hss}[HSS]{Home Subscriber Server}
\acro{hw}[HW]{Hardware}
\acro{id}[ID]{Identification}
\acro{iscp}[ISCP]{Interference Signal Code Power}
\acro{itu}[ITU]{International Telecommunication Union}
\acro{kdd}[KDD]{Knowledge Discovery in Databases}
\acro{km}[KM]{Knowledge Mining}
\acro{knn}[K-NN]{K-Nearest Neighbors}
\acro{kpi}[KPI]{Key Performance Indicator}
\acro{kqi}[KQI]{Key Quality Indicator}
\acro{lof}[LOF]{Local Oulier Factor}
\acro{lte}[LTE]{Long Term Evolution}
\acro{ltea}[LTE-A]{Long Term Evolution Advanced}
\acro{mac}[MAC]{Medium Access Control}
\acro{mca}[MCA]{Minor Component Analysis}
\acro{mdt}[MDT]{Minimization of Drive Tests}
\acro{mln}[MLN]{Markov Logic Networks}
\acro{mme}[MME]{Mobility Management Entity}
\acro{mos}[MOS]{Mean Opinion Score}
\acro{mro}[MRO]{Mobility Robustness Optimization}
\acro{ne}[NE]{Network Element}
\acro{ngmn}[NGMN]{Next Generation Mobile Networks}
\acro{nm}[NM]{Network Management}
\acro{ns3}[ns-3]{Network Simulator 3}
\acro{oam}[OAM]{Operations, Administration, and Maintenance}
\acro{ofd}[OFD]{Operational Fault Detection}
\acro{ofdm}[OFDM]{Orthogonal Frequency-Division Multiplexing}
\acro{optics}[OPTICS]{Ordering points to identify the clustering structure}
\acro{oss}[OSS]{Operations Support System}
\acro{pca}[PCA]{Principal Component Analysis}
\acro{pci}[PCI]{Physical Cell Identity}
\acro{pesq}[PESQ]{Perceptual Evaluation of Speech Quality}
\acro{pgw}[PGW]{Packet Gateway}
\acro{phr}[PHR]{Power Headroom}
\acro{pi}[PI]{Performance Indicator}
\acro{pm}[PM]{Performance Monitoring}
\acro{prach}[PRACH]{Physical Random Access Channel}
\acro{qoe}[QoE]{Quality of Experience}
\acro{qos}[QoS]{Quality of Service}
\acro{qpm}[QPM]{Quality and Performance Management}
\acro{ra}[RA]{Recovery Analysis}
\acro{rach}[RACH]{Random Access Channel}
\acro{ran}[RAN]{Radio Access Network}
\acro{rat}[RAT]{Radio Access Technology}
\acro{rcef}[RCEF]{RRC Connection Establishment Failure}
\acro{rf}[RF]{Radio Frequency}
\acro{rlf}[RLF]{Radio Link Failure}
\acro{rnc}[RNC]{Radio Network Controller}
\acro{roc}[ROC]{Receiver Operating Characteristic}
\acro{rrc}[RRC]{Radio Resource Control}
\acro{rrm}[RRM]{Radio Resource Management}
\acro{rscp}[RSCP]{Received Signal Code Power}
\acro{rsrp}[RSRP]{Reference Signal Received Power}
\acro{rsrq}[RSRQ]{Reference Signal Received Quality}
\acro{rssi}[RSSI]{Received Signal Strength Indicator}
\acro{sc}[SC]{Sleeping Cell}
\acro{sdcch}[SDCCH]{Stand-alone Dedicated Control Channel}
\acro{sgsn}[SGSN]{Serving GPRS Support Node}
\acro{sgw}[SGW]{Serving Gateway}
\acro{sinr}[SINR]{Signal to Interference plus Noise Ratio}
\acro{sir}[SIR]{Signal to Interference Ratio}
\acro{socrates}[SOCRATES]{Self-Optimisation and self-ConfiguRATion in wirelEss networkS}
\acro{som}[SOM]{Self-Organizing Maps}
\acro{son}[SON]{Self-Organizing Network}
\acro{sorte}[SORTE]{Second ORder sTatistic of the Eigenvalues}
\acro{sqm}[SQM]{Service Quality Management}
\acro{sting}[STING]{STatistical INformation Grid}
\acro{svd}[SVD]{Singular Value Decomposition}
\acro{svm}[SVM]{Support Vector Machine}
\acro{sw}[SW]{Software}
\acro{tce}[TCE]{TRACE Collection Entity}
\acro{tnr}[TNR]{True Negative Rate}
\acro{ttt}[TTT]{Time to Trigger}
\acro{ue}[UE]{User Equipment}
\acro{ul}[UL]{Uplink}
\acro{umts}[UMTS]{Universal Mobile Telecommunications System}
\acro{utra}[UTRA]{Universal Terrestrial Radio Access}
\acro{utran}[UTRAN]{Universal Terrestrial Radio Access}
\acro{wcdma}[WCDMA]{Wideband Code Division Multiple Access}
\acro{wifi}[WiFi]{Wireless Fidelity}
\end{acronym}
\maketitle
\begin{abstract}
This article presents an automatic malfunction detection framework based on data mining approach to analysis of network event sequences. The considered environment is \ac{lte} for \ac{umts} with sleeping cell caused by random access channel failure. Sleeping cell problem means unavailability of network service without triggered alarm. The proposed detection 
framework uses N-gram analysis for identification of abnormal behavior in sequences of network events. 
These events are collected with \ac{mdt} functionality standardized in \ac{lte}. Further processing applies dimensionality reduction, anomaly detection with \ac{knn}, cross-validation, post-processing 
techniques and efficiency evaluation. Different anomaly detection approaches proposed in this paper are compared against each other with both classic data mining metrics, such as F-score and \ac{roc} curves, and a newly proposed heuristic approach. 
Achieved results demonstrate that the suggested method can be used in modern performance monitoring systems for 
reliable, timely and automatic detection of random access channel sleeping cells.
\end{abstract}


\section{Introduction}
\label{sec:introduction}
Modern cellular mobile networks are becoming increasingly diverse and complex, due to coexistence of multiple \acp{rat}, and their corresponding releases. Additionally, small cells are actively deployed to complement the macro layer coverage, and this trend will only grow. In the future this situation is going to evolve towards even higher complexity, as in 5G networks there will be much more end-user devices, served by different technologies, and connected to cells of different types. New applications and user behavior patterns are daily coming into play. In such environment network performance and robustness are becoming critical values for mobile operators. In order to achieve these goals, efficient flow of \ac{qpm} \cite{Hamalainen12}, which is a sequence of fault detection, diagnosis and healing, should be developed and applied in the network in addition to other optimization functions. 

Concept of \ac{son} \cite{NGMN_req08_jr,NGMN_cases08_jr} has been proposed to automate and optimize the most tedious manual tasks in mobile networks, including \ac{qpm}. Automation is the key idea in \ac{son} and it has been proposed for self-configuration, self-optimization and self-healing in \ac{lte} and \ac{umts} networks \cite{3GPP_son_jr, Hamalainen12, SocratesD21_jr}. 
In traditional systems detection, diagnosis and recovery of network failures is mostly manual task, and it is heavily based on pre-defined thresholds, aggregation and averaging of large amounts of performance data -- so called \acp{kpi}. Self-healing \cite{Ramiro12}, \cite{32.541_jr} automates the functions of \ac{qpm} process to improve reliability of network operation. Though, self-healing is still among the least studied functions of \ac{son} at the moment, and the developed solutions and use cases require improvement prior to application in the real networks. This is especially important for non-trivial network failures such as sleeping cell problem \cite{Cheung06_patent_jr, Cheung05, Hamalainen12}. This is a special term used to denote a breakdown, which causes partial or complete degradation of network performance, and which is hard to detect with conventional \ac{qpm} within reasonable time. Thus, in the research and standardization community automatic fault detection and diagnosis functions, enhanced with the most recent advancements in data analysis, are seen as the future of self-healing. 
Thus, development of improved self-healing functions for detection of sleeping cell problems, through application of anomaly detection techniques is of high importance nowadays. This article presents a novel framework based on N-gram analysis of \ac{mdt} event sequences for detection of random access channel sleeping cells. 

The rest of this paper is organized as follows. Section \ref{sec:qpm} describes common practices of quality and performance management in mobile networks, including \ac{mdt} functionality, and advanced methods based on knowledge mining algorithms. Section \ref{sec:sleepcell} defines the concept of sleeping cell and its possible root cause failures. In Section \ref{sec:experim_setup} simulation environment, 
assumptions and random access channel problem are presented. Also Section \ref{sec:experim_setup} describes the generated and analyzed performance \ac{mdt} data. Section \ref{sec:sc_framework} concentrates on the suggested sleeping cell detection knowledge mining framework. It includes overview of the applied anomaly detection methods: \ac{knn} anomaly 
outlier scores, N-gram, minor component analyses, post-processing and data mining
performance evaluation techniques. Section \ref{sec:results} is devoted to the 
actual research results. Data structures at different stages of 
analysis are shown, and efficiency of different post-processing 
methods is compared. In Section \ref{sec:conclusions} the concluding remarks regarding the findings of the presented research are given.
\section{Quality and Performance Management in Cellular Mobile Networks}
\label{sec:qpm}
Performance management in wireless networks includes three main 
components: data collection, analysis and results interpretation. Data 
gathering can be done either by aggregation of cell-level statistics - 
collection of \acp{kpi}, or collection of detailed performance data with drive tests. The main weaknesses in analysis of \acp{kpi} are that a lot of statistics is left out at the aggregation stage, due to averaging over time, element and because fixed threshold values are applied. Even thought drive test campaigns provide far more elaborate information regarding network performance, they are expensive to carry out and do not cover overall area of network operation. Root cause 
analysis is done manually in majority of cases, and because of that there is a 
room for more intelligent approaches to detection and diagnosis of 
network failures, e.g. with data mining and anomaly detection 
techniques. This would provide possibility to automate performance 
monitoring task furthermore.
\subsection{Minimization of Drive Tests}
\label{subsec:mdt}
Yet another way to improve network \ac{qpm} is to collect a detailed performance database. This is enabled with \ac{mdt} functionality standardized in \ac{3gpp}  \cite{36.805_jr}. \ac{mdt} is designed for 
automatic collection and reporting of user measurements, where possible complemented with location information. Collected data is then reported to the 
serving cell, which in turn sends it to \ac{mdt} server 
\cite{MDT_paper12_part1}. Thus, large amount of network and user performance is available for analysis. This is where the power of data mining and anomaly detection can be applied. 

Specification describes several use cases for \ac{mdt}: 
improvement of network coverage, capacity, mobility robustness and end 
user quality of service \cite{Hamalainen12}. According to the 
standard, \ac{mdt} measurements and reporting can be done both in idle and connected \ac{rrc} modes. In logged \ac{mdt}, \ac{ue} stores 
measurements in memory, and reporting is done at the next 
transition from idle to connected state. In immediate \ac{mdt}, measurements are reported as soon as they are done through existing connection. 
In turn, there are two measurement modes in immediate \ac{mdt}: periodic and event-triggered \cite{MDT_paper12_part1}. Periodic measurements are very useful for 
initial network deployment coverage and capacity verification as they 
provide detailed map of network performance, say in terms of signal 
propagation or throughput. The main disadvantage of periodic 
measurements is that they consume a lot of network and user resources. In contrast, event-triggered approach provides less information regarding the 
network status, but can be very efficient for mobility robustness and 
resource savings. In our study, immediate event-triggered \ac{mdt} is used for collection of performance database. Table \ref{tab:mdt_events} presents the list of  
network events which triggered \ac{mdt} measurements and reporting.
\begin{table}
\caption{Network events triggering \ac{mdt} measurements and reporting}
\label{tab:mdt_events} \centering
\begin{tabular}{l}
\hline \hline
PL PROBLEM - Physical Layer Problem \cite{36.331_jr}. \\
RLF - Radio Link Failure \cite{sesia2011lte}.\\
RLF REESTAB. - Connection reestablishment after RLF. \\
A2 RSRP ENTER - RSRP goes under A2 enter threshold. \\
A2 RSRP LEAVE - RSRP goes over A2 leave threshold.  \\
A2 RSRQ ENTER - RSRQ goes over A2 enter threshold. \\
A3 RSRP - A3 event, according to 3GPP specification. \\
HO COMMAND - handover command received \cite{sesia2011lte}.\\
HO COMPLETE - handover complete received \cite{sesia2011lte}.\\
\hline
\end{tabular}
\end{table}

\subsubsection{Location Estimation in \ac{mdt}}
\label{subsec:location_estimation}
One of the important features of \ac{mdt} is collection of geo-location 
information at the measurement time moments. Whenever \ac{ue} location is provided in \ac{mdt} report there are several ways 
to associated it with particular cell, such as: serving cell ID, dominance maps and a new approach based on target cell ID information. 

\begin{figure}
\centering
\includegraphics[scale=0.52]{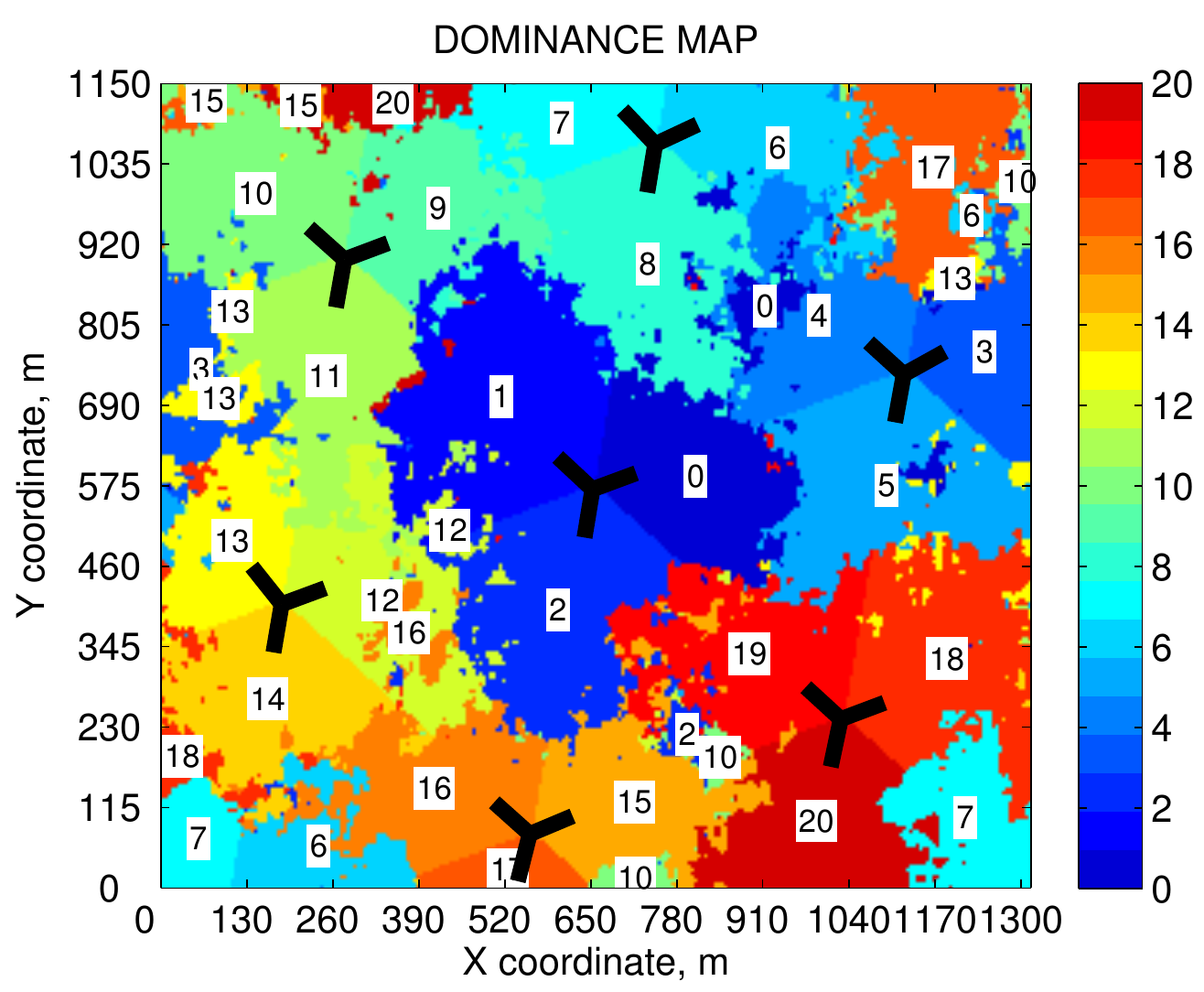}
\caption{Wrap around Macro 21 slow faded dominance map}
\label{fig:scenario}
\end{figure}

Serving cell ID is available with \ac{mdt} event-triggered report, even for 
early releases of \ac{lte}. However, in case of coverage hole or problems 
with new connection establishment, this approach can lead to mistakes 
in \ac{ue} location association, because the faulty cell would never become serving in the worst case scenario. This limits the usage of serving cell method for sleeping cell detection. To overcome the problem presented above, a dominance maps method can be used. This is a map, which demonstrates the  
\ac{enb}\footnote{\ac{eutran}} with dominating, i.e. strongest radio signal in each point of the network, see Fig. \ref{fig:scenario}. Creation of dominance map requires information about path loss and slow fading.
The main advantage of dominance maps is that 
mapping of cell ID to location coordinate of \ac{ue} \ac{mdt} measurement is 
very precise, and this results 
in higher accuracy of sleeping cell detection. The downside dominance maps approach is that it requires a lot of 
detailed input measurement information. Though, \ac{mdt} functionality is one of the ways to create such maps fast and relatively simple. Additionally, 
more accurate user location information is going to be available with 
deployment of newer releases of mobile networks \cite{E911_jr}.

The last method for cell ID and \ac{ue} report location association uses 
target cell ID feature. The main advantage of this approach is that it 
does not require serving cell ID, user geo-positioning location or knowledge about 
network dominance areas. This eases the requirements for \ac{mdt} data 
collection in amount of details regarding user location. The problem 
of mapping on the basis of target cell ID, is that it might be useful 
for detection of only particular types of network problem, such random 
access \ac{sc}. Efficiency of this method for detection of other 
malfunctions is subject for further verification.

The key aspects which should be taken into account when selecting a location association method are accuracy and amount of information to create mapping 
between cell and user location.

\subsection{Advanced data analysis approaches in \ac{qpm}}
\label{sec:adv_qpm}
Studies in advanced data analysis for \ac{qpm} can be divided to several groups. In certain studies, the data reported by the users is used for the analysis. For instance, in \cite{Mueller08} authors suggest a method for detection 
of sleeping cells, caused by transmitted signal strength problem, on 
the basis of neighbor cell list information. Application of non-trivial pre-
processing and different classification algorithms allowed to achieve relatively good accuracy in detection of cell hardware faults. However, the 
proposed anomaly detection system is prone to have relatively high 
false rate. In \cite{Turkka11} a method based on analysis of TRACE-based user data with diffusion maps is presented. More extensive application of diffusion maps for network performance monitoring can also be found in \cite{Kassis10}.

Even though, user level statistics is more detailed, still majority of studies devoted to improvement of \ac{qpm} rely on cell-level data. The first proposals of sleeping cell detection automation using statistical methods of network monitoring are presented in \cite{Cheung05, Cheung06_patent_jr}. Preparation of normal cell load profile and evaluation of the deviation in observed cell behavior is suggested as a way for identification of problematic cells. The idea of statistical approach has been further studied in \cite{Szilagyi11}, \cite{Szilagyi12}, \cite{Novaczki13}, where a profile-based system for fault detection and diagnosis is proposed. Bayesian networks have also been applied for diagnosis and root cause probability estimation, given certain \acp{kpi} \cite{Khanafer08, Barco08, Barco09, Barco10}. The complications here are  preparation of correct probability model and appropriate \ac{kpi} threshold parameters. More advanced data mining methods are applied to analysis of cell-level performance statistics, and novel ensemble methods of classification algorithms is proposed \cite{Ciocarlie13, Ciocarlie14}. In \cite{Ciocarlie14new_framework, Ciocarlie14managing} application of classification and clustering methods for detection and diagnosis of strangely behaving network regions is presented. Some studies also consider neural network algorithms for detection of malfunctions \cite{Raivio03, Laiho05_som}.

The largest drawback of processing cell level data is that collection of appropriate 
statistical base takes substantial amount of time, and can vary from days to months.  This increases reaction time in case of outages and does not completely solve the problems of operators in optimization of their \ac{qpm}. In order to overcome weaknesses of analysis based on cell \acp{kpi}, our studies are  concentrated at the analysis of the user-level data, collected with immediate \ac{mdt} functionality \cite{Hapsari12, Johansson12}. In the early works cell outage detection 
caused by signal strength problems (antenna gain failure) is studied \cite{Chernogorov10, ChernogorovVTC11, Jtu_journal12}. 
This area matches the \ac{3gpp} use case called ``cell outage detection'' \cite{32.541_jr}. Identification of the cell, in malfunction condition is done by means of analysis of numerical properties of multidimensional dataset. Each data point represents either periodic or event-triggered user measurement. Such methods as diffusion maps dimensionality reduction algorithm, k-means clustering and k-nearest neighbor classification methods are applied. 

To increase robustness of the proposed solutions in \ac{mdt} data analysis and make the developed detection system suitable for application in real networks, a more sophisticated experimental setup is considered. Sleeping cell caused by malfunction of random access channel, discussed in Section \ref{sec:sleepcell}, does not produce coverage holes from perspective of radio signal, but still makes service unavailable to the subscribers. This problem is considered to be one of the most complex for mobile network operators, as detection of such failures may take days or even weeks, and negatively affects user experience \cite{Hamalainen12}. To make fault detection framework more flexible and independent from user behavior, such as variable mobility and traffic variation, analysis of numerical characteristics of \ac{mdt} data is substituted with processing of \textit{network event sequences} with N-gram method. Network events can include different mobility or signaling related nature, such as A2, A3 or handover complete message \cite{Holma11}. Initial results in this area are presented in \cite{Chernogorov13}.
\section{Sleeping Cell Problem}
\label{sec:sleepcell}
Sleeping cell is a special kind of cell service degradation. It means 
malfunction resulting in network performance decrease, invisible for a 
network operator, but affecting user \ac{qoe}. On one hand, detection of 
sleeping cell problem with traditional monitoring systems is 
complicated, as in many cases \ac{kpi} thresholds do not indicate the 
problem. On the other hand fault identification can be very sluggish, 
as creation of cell behavior profile requires long time, as it is discussed in the previous section. Regular, less sophisticated 
types of failures usually produce cell level alarms to performance 
monitoring system of mobile network operator. In contrast, for  
sleeping cells degradation occurs seamlessly and no direct 
notification is given to the service provider.

In general, any cell can be called degraded in case if it is not 100\% 
functional, i.e. its services are suffering in terms of quality, what 
in turn affects user experience. There are 3 distinguished extents of cell performance degradation Classification of sleeping cells, 
depending on the extent of performance degradation from the lightest, to the most severe \cite{Cheung05},\cite{Cheung06_patent}: \textit{impared} or \textit{deteriorated} - smallest negative impact on the provided service, \textit{crippled} - characterized by a severely decreased capacity, and \textit{catatonic} - kind of outage which 
leads to complete absence of service in the faulty area, such cell does 
not carry any traffic.

Degradation can be caused by malfunction of different hardware or software components of 
the network. Depending on the failure type, different extent of 
performance degradation can be induced. In this study the considered 
sleeping cell problem is caused by \ac{rach} failure. 
This kind of problem can appear due to \ac{rach} misconfiguration, 
excessive load or software/firmware problem at the \ac{enb} side 
\cite{Amirjoo09_rach}, \cite{Yilmaz11}. \ac{rach} malfunction leads to inability 
of the affected cell to serve any new users, while earlier connected 
\acp{ue} still get served, as pilot signals are transmitted. This problem 
can be classified to crippled sleeping cell type, and with time it 
tends to become catatonic. In many cases \ac{rach} 
problem becomes visible for the operator only after a long observation time or even due to user complains. For this reason, it is very important to timely detect such cells and apply recovery actions.

\subsubsection{Random Access Sleeping Cell}
\label{subsub:sleepcell_model}
Malfunction of \ac{rach} can lead to 
severe problems in network operation as it is used for connection 
establishment in the beginning of a call, during handover to another 
cell, connection re-establishment after handover failure or \ac{rlf} 
\cite{sesia2011lte}. Malfunction of random access in cell with ID 1, is caused by erroneous behavior of T304 timer \cite{36.331_jr}, which expires before random access procedure is finished. Thus, whenever \ac{ue} tries to initiate random access to cell 1, this attempt fails. Malfunction area covers around 5 \% of the overall network. \\

\section{Experimental Setup}
\label{sec:experim_setup}
\subsection{Simulation environment}
\label{subsec:simulation}
Experimental environment is dynamic system level simulator of \ac{lte} network, designed according to \ac{3gpp} Releases 8, 9, 10 and partly 11. Throughput, spectral efficiency and mobility-related behavior of this simulator is validated against results from other simulators of several companies in \ac{3gpp} \cite{s36.839, Kolehmainen07_msc, Kela07_msc}. Step resolution 
of the simulator is one \ac{ofdm} symbol. Methodology for mapping link level SINR to the system level is presented in \cite{Brueninghaus05}. Simulation 
scenario is an improved 3GPP macro case 1 \cite{36.814_jr} with wrap-around layout, 21 cells (7 base stations with 3-sector antennas), and inter-site distance of 500 meters. Modeling of propagation 
and radio link conditions includes slow and fast fading. 
Users are spread randomly around the network, so that on average there 
are 15 dynamically moving \acp{ue} per cell. The main configuration parameters of the simulated network are shown in Table \ref{tab:simulation_param}.
\begin{table*}[!t]
\renewcommand{\arraystretch}{1.3}
\caption{General Simulation Configuration Parameters}
\label{tab:simulation_param}
\centering
\begin{tabular}{p{2.5cm}|p{2.5cm}||p{2.5cm}|p{2.5cm}}
\hline
\bfseries Parameter & \bfseries Value & \bfseries Parameter & \bfseries Value\\
\hline\hline
Cellular layout & Macro 21 Wrap-around & Number of cells & 21 \\
\acp{ue} per cell & 17 &
Inter-Site Distance & 500 m \\
Link direction & Downlink &
RRC IDLE mode & Disabled \\
User distribution in the network & Uniform &
Maximum BS TX power & 46 dBm \\
Initial cell selection criterion & Strongest RSRP value&
Handover margin (A3 margin) & 3 dB \\
Handover time to trigger & 256 ms&
Hybrid Adaptive Repeat and reQuest (HARQ) & Enabled \\
Slow fading standard deviation & 8 dB &
Slow fading resolution & 5 m \\
Simulation length & 572 s (~9.5 min)&
Simulation resolution & 1 time step = 71.43 $\mu s$\\
Network synchronicity mode & Asynchronous &
Max number of \acp{ue}/cell & 20 \\
\ac{ue} velocity & 30 km/h &
Duration of calls &  \\
Traffic model & Constant Bit Rate 320 kbps &
Normal and Reference cases & Simulation without sleeping cell \\
Problematic case & Simulation with \ac{rach} problem in cell 1& &\\
A2 RSRP Threshold & -110 &
A2 RSRP Hysteresis& 3 \\
A2 RSRQ Threshold& -10 &
A2 RSRQ Hysteresis& 2 \\
\hline
\end{tabular}
\end{table*}

\subsection{Generated Performance Data}
\label{subsec:input_data}
Generated performance data includes dominance map information and \ac{mdt} log, which contains the following fields:
\begin{itemize}
\item \ac{mdt} triggering event ID. The list of possible events is presented in Table \ref{tab:mdt_events}. This is a categorical (nominal) and sequential data, i.e. 
sequences of events are meaningful from data mining perspective;
\item \ac{ue} ID. This is also categorical data;
\item \ac{ue} location coordinates [m]. It is numerical, spatial data;
\item Serving and target cell ID -- spatial, categorical data.
\end{itemize}
It is important to know the type of the analyzed data to construct efficient knowledge mining framework \cite{Chandola09, Han06}.

Simulations done for this study cover three types of network behavior:
``normal'' -- network operation \textit{without} random access sleeping cell;
``problematic'' -- network \textit{with} \ac{rach} failure in cell 1; ``reference'' --  \textit{no sleeping cell}, but different slow and fast fading maps, i.e. if compared to ``normal'' case, propagation-wise it is a different network. The latter case is used for validation purposes. All three of these cases have different mobility random seeds, i.e. call start locations and \ac{ue} traveling paths are not the same.
Each of the cases are represented with 6 data chunks. The training and testing phases of sleeping cell detection are done with pairs of \ac{mdt} logs by means of K-fold approach \cite{Han06}. For example, ``normal''-``problematic'', or ``normal''-``reference'' cases are considered. Thus, in total there are 72 unique combinations of analyzed \ac{mdt} log pairs, which is rather statistically reliable data base. 

\section{Sleeping Cell Detection Framework}
\label{sec:sc_framework}
\begin{figure}
\centering
\includegraphics[scale=0.435]{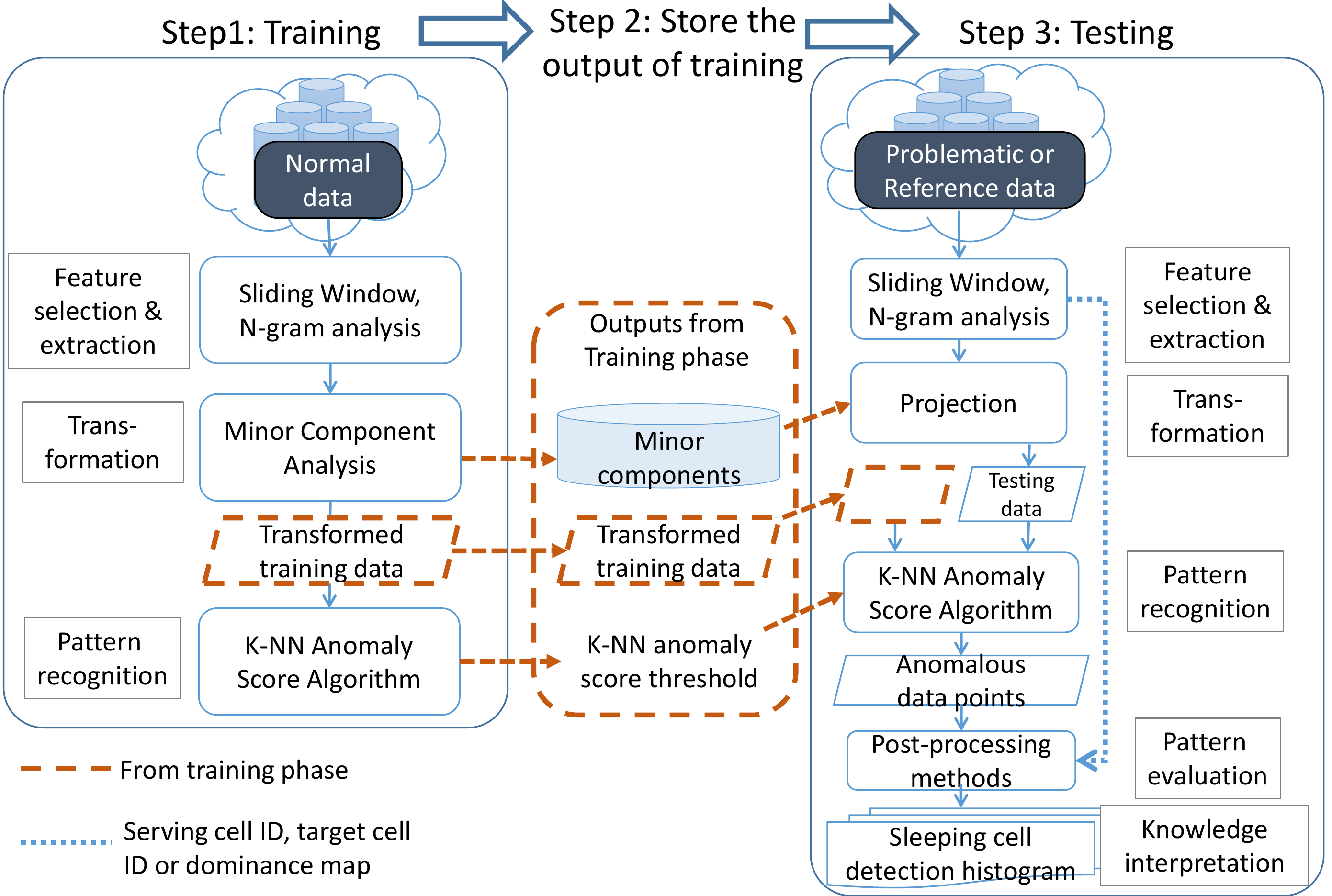}
\caption{Sleeping Cell Detection Framework}
\label{fig:sc_framework}
\end{figure}

The core of the presented study is sleeping cell detection framework based on knowledge mining, Fig. \ref{fig:sc_framework}. Both training and testing phases are done in accordance to the process of \ac{kdd}, which includes the following steps \cite{Fayyad96}, \cite{Han06}: data cleaning, integration from 
different sources, feature selection and extraction, transformation, 
pattern recognition, pattern evaluation and knowledge presentation. The constructed  data analysis framework for sleeping cell detection is semi-supervised,
because unlabeled error-free data is used for training of the data mining algorithms. 
In testing phase problematic data is analyzed to detect abnormal behavior. Reference data is used for testing in order to verify how much the designed framework is prone to make false alarms. 
\subsection{Feature Selection and Extraction}
\label{subsec:preprocessing}
Feature selection and extraction is the first step of sleeping cell detection. At this stage, input data is prepared for further analysis. Pre-processing is needed as reported \acp{ue} \ac{mdt} event sequences have variable lengths, depending on the user call duration, velocity, traffic distribution and network layout. 
\subsubsection{Sliding Window Pre-processing}
\label{subsub:sliding_window}
Sliding window approach \cite{Rabin10} allows to divide calls to \textit{sub-calls} of constant length, an by that to unify input data. 
There are two parameters in sliding window algorithm: window 
size $m$ and step $n$. After transformation, one sequence of $N$ events (a call) is represented by 
several overlapping (in case if $n < m$) sequences of equal sizes, except for the last sub-call, which is the remainder from $N$ modulo $n$.

In the presented results overlapping sliding window size is 15, and the step is 10 events. Such setup allows to maintain the context of the data 
after processing \cite{Kassis10}. 
The number of calls and sub-calls for all three data sets are shown in Table \ref{tab:calls_sub_calls}. 
\begin{table}
\caption{Number of calls and sub-calls in analyzed data}
\label{tab:calls_sub_calls}
\centering
\begin{tabular}{l|c|c|c}
\hline
\textbf{Amount / Dataset}& \textbf{Normal}& \textbf{Problem}&\textbf{Reference }\\
\hline
\hline
Calls (all)        & 2530 & 1940 & 2540 \\
Sub-calls (all)    & 7230 & 7134 & 7201 \\
Normal sub-calls   & 6869 &	5932 & 6821 \\
Abnormal sub-calls & 361  & 1202 & 380  \\
\hline
\end{tabular}
\end{table}
\subsubsection{N-Gram Analysis}
\label{subsub:n_gram}
When input user-specific \ac{mdt} log entries are standardized with sliding window method, the data is transformed from sequential to numeric format. It is done with N-gram analysis method , widely used e.g. for natural language 
processing and text analysis applications such as speech recognition, 
parsing, spelling, etc. \cite{Brown92, Nagao94anew, Cavnar94, Haidar2012, Islam2009}. In addition, N-gram is applied for whole-genome protein sequences \cite{Ganapathiraju_2002} and for computer virus detection \cite{ChoiKCK11, David09}.

N-gram is a sub-sequence of N overlapping items or units from a given 
original sequence. The items can be characters, letters, words or 
anything else. The idea of the method is to count how many times each sub-sequence occurs. This is the transformation from sequential to numerical space.

Here is an example of \textit{N}-gram analysis application for two words: `performance' and `performer', $N=2$, and a single unit is a character. The resulting frequency matrix after $N$-gram processing is shown in Table \ref{tab:km_n_gram}.
\begin{table}
\caption{Example of \textit{N}-gram analysis per character, $N=2$.}
\label{tab:km_n_gram} \centering
\begin{tabular}{|l|c|c|c|c|c|c|c|c|c|c|}
\hline 
Analyzed word & pe & er & rf & or & rm & ma & me & an & nc & ce \\ \hline
performance     & 1 & 1 & 1 & 1 & 1 & 1 & 0 & 1 & 1 & 1\\ \hline 
performer       & 1 & 2 & 1 & 1 & 1 & 0 & 1 & 0 & 0 & 0\\ \hline
\end{tabular}
\end{table}

\subsection{Dimensionality Reduction with Minor Component Analysis}
\label{subsub:mca}
Dimensionality reduction is applied to convert high-
dimensional data to a smaller set of derived variables. In the presented study \ac{mca} method is applied \cite{Luo97}. This algorithm has been selected selected on the basis of comparison with other 
dimensionality reduction methods such as \ac{pca} \cite{Jolliffe02} and diffusion maps \cite{Coifman06}. \ac{mca} extracts components of covariance matrix of the input data set and uses minor components (eigenvectors with the smallest eigenvalues of covariance matrix). 6 minor components are used as a basis of the embedded space. This number is defined by means of \ac{sorte} method \cite{sorte09, sorte10}. 

\subsection{Pattern Recognition: K-NN Anomaly Score Ourlier Detection}
\label{subsec:data_mining}
In order to extract abnormal instances from the testing dataset \ac{knn} anomaly outlier score algorithm is applied. In contrast with \ac{knn} classification, method is not supervised, but semi-
supervised, as the training data does not contain any abnormal labels. 
In general, there are two approaches concerning the implementation of 
this algorithm; anomaly score assigned to each point is either the sum 
of distances to k nearest neighbors \cite{Angiulli2002} or distance to 
k-th neighbor \cite{Ramaswamy2000}. The first method is employed in the presented sleeping cell detection framework, as it is more statistically robust. Thus, the algorithm assigns an anomaly score to every sample in the analyzed data based on the sum of distances to k nearest 
neighbors in the embedded space. Euclidean metric is applied as similarity measure. Points with the largest anomaly scores are called outliers. Separation to normal and abnormal classes is defined by threshold parameter $T$, equal to $95^{th}$ percentile of anomaly scores in the training data. 

Configuration parameters of data analysis algorithms in the presented sleeping cell detection framework are summarized in Table \ref{tab:ad_conf_param}.
\begin{table}
\caption{Parameters of algorithms in sleeping cell detection framework}
\label{tab:ad_conf_param} \centering
\begin{tabular}{l||l}
\hline \hline
\textbf{Parameter} & \textbf{Value} \\ 
\hline
Number of chunks in K-fold method per dataset & 6 \\

Sliding window size & 15 \\

Sliding window step & 10 \\

$N$ in N-gram algorithm & 2 \\

Number of nearest neighbors ($k$) in \ac{knn} algorithm & 35 \\

Number of minor components & 6 \\
\hline
\end{tabular}
\end{table}

\subsection{Pattern Evaluation}
\label{subsec:post_processing}
The main goal of pattern evaluation is conversion of output information from \ac{knn} anomaly score algorithm to knowledge about location of the network malfunction, i.e. \ac{rach} sleeping cell. This is achieved with post-processing of the anomalous data samples through analysis of their correspondence to particular network elements, such as \acp{ue} and cells. 4 post-processing methods are developed for this purpose. The essence of these methods, discussed throughout this section, is  
reflected in their names. The first part describes which geo-location 
information is used for mapping data samples to cells, e.g. dominance map information, target or serving cell ID. The second part denotes what is used as feature space for post-processing. It can be either ``sub-calls'', when rows of the dataset are used as features or ``2-gram'', when individual event pair combinations, i.e. columns of the dataset are used as features. The last, third part of the method name describes is analysis considers the difference between training and testing data (``deviation'' keyword), or whether only information about testing set is used to build sleeping cell detection histogram. 

Output from the post-processing methods described above is a set of 
values - sleeping cell scores, which correspond to each cell in the analyzed network. High value of this score means higher abnormality, and hence probability of failure. To achieve 
clearer indication of problematic cell presence, additional non-linear 
transformation is applied. It is called amplification, as it allows to emphasize problematic areas in the sleeping cell histogram. Sleeping cell score of each cell is divided 
by the sum of \ac{sc} scores of all non-neighboring cells. Sleeping cell scores, received after post-processing and amplification 
are then normalized by the cumulative \ac{sc} score of all cells 
in the network. Normalization is necessary to get rid of dependency on 
the size of the dataset, i.e. number of calls and users.

\subsection{Knowledge Interpretation and Presentation}
\label{subsec:know_present}
The final step of the data analysis framework is visualization of the fault detection results. It is done with construction of a sleeping cell detection histogram and network heat map. 
However, sleeping cell histogram does not show how cells are related to each other: are they neighbors or not, and which area of the network is causing problems. Heat map method shows more anomalous network regions with darker and larger spots, while normally operating regions are in light grey color. The main benefit of network heat map is that mobile network topology and neighbor relations between cells are illustrated.

\subsubsection{Performance Evaluation}
\label{subsub:perf_meas}
To apply data mining performance evaluation metrics labels of data points must be know. Cell is labeled as abnormal if its \ac{sc} score deviates more than 3$\sigma$ (standard deviation of sleeping cell scores) from the mean \ac{sc} of score in the network. Mean value and standard deviation of the sleeping 
cell scores are calculated altogether from 72 runs produced by K-fold method for ``normal''-``problematic'', and ``normal''-``reference'' dataset pairs.  Availability of the labels and the outcomes of different post-processing 
methods enables application of such data mining performance metrics as accuracy, 
precision, recall, F-score, \ac{tnr} and \ac{fpr} \cite{Guillet07}. On the basis of these scores \ac{roc} curves are plotted.

In addition to the conventional performance evaluation metrics described above, a heuristic method is applied to complement the analysis. This approach measures how far is the achieved performance from the \textit{a priori} known ideal solution.
Performance of the sleeping cell detection algorithm can be 
described by a point in the space \textit{``sleeping cell 
magnitude''}-\textit{``cumulative standard deviation''}. \textit{``Sleeping cell magnitude''} is the highest sleeping cell score, and a sum of all sleeping cell scores is \textit{``cumulative standard deviation''}.
This plane contains two points of interest: in case of malfunctioning network, the ideal sleeping cell detection algorithm would have coordinate $[0; 100]$. In case of error-free network, the 
ideal performance is point $[0; 100/N_{\text{cells in the network}}]$.
Thus, the smaller the Euclidean distance between the achieved and ideal sleeping cell histograms, the better the performance of the sleeping cell detection algorithm.
\section{Results of Sleeping Cell Detection}
\label{sec:results}

This section presents the results of sleeping cell detection for different post-processing algorithms. In addition, the data at different stages of the detection process is illustrated. Then performance metrics are used to compare effectiveness of the developed \ac{sc} identification algorithms.

\subsection{Pre-processing and K-NN Anomaly Score Calculations}
\label{subsec:res_preproc_and_scores}
\begin{figure}
\centering
\subfloat[Normal training dataset in the embedded space.]{\label{fig:embedded_train_norm}\includegraphics[scale=0.42]{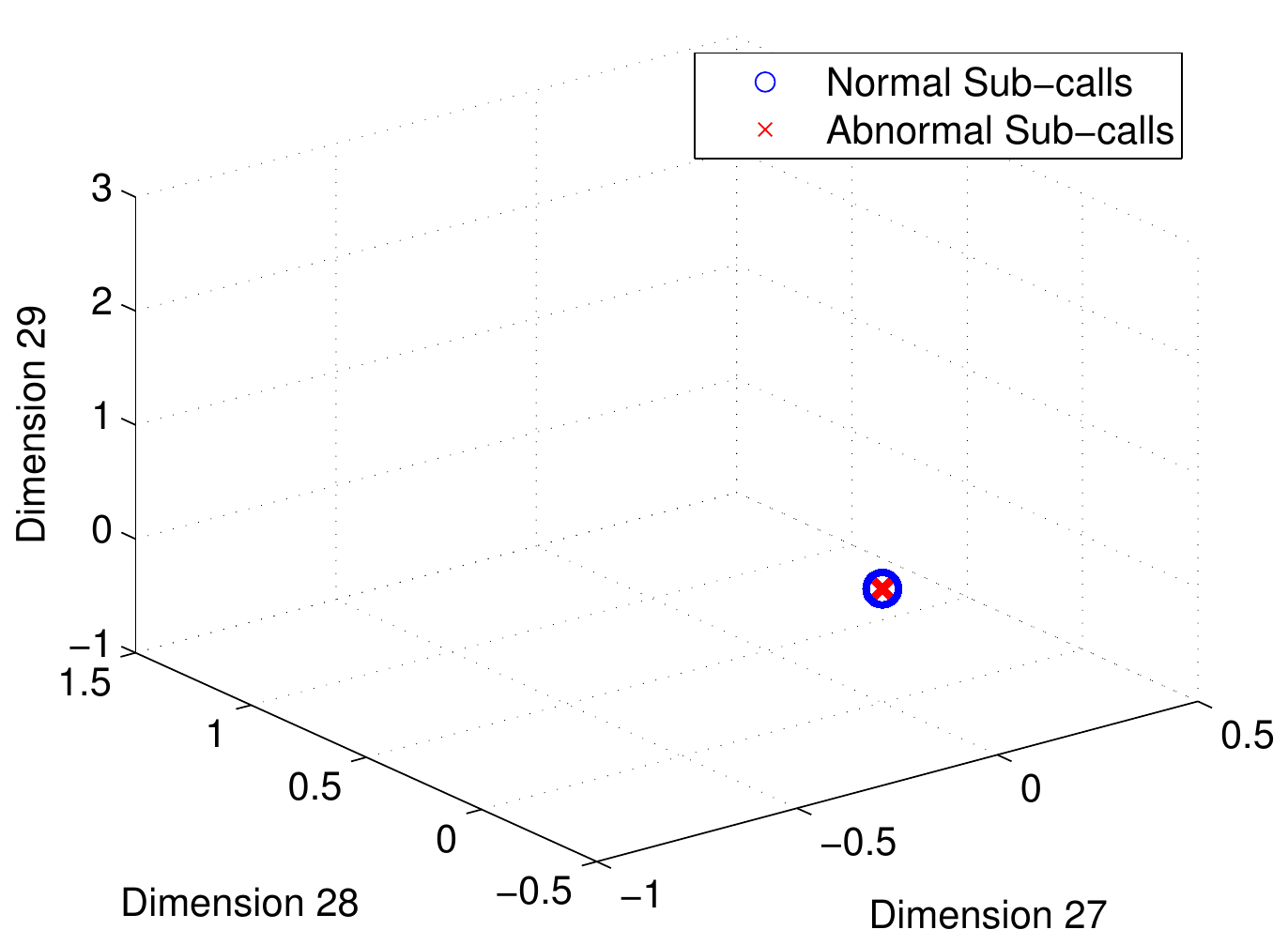}}
\subfloat[Sorted outlier scores of normal training dataset.]{\label{fig:knn_scores_norm}\includegraphics[scale=0.4]{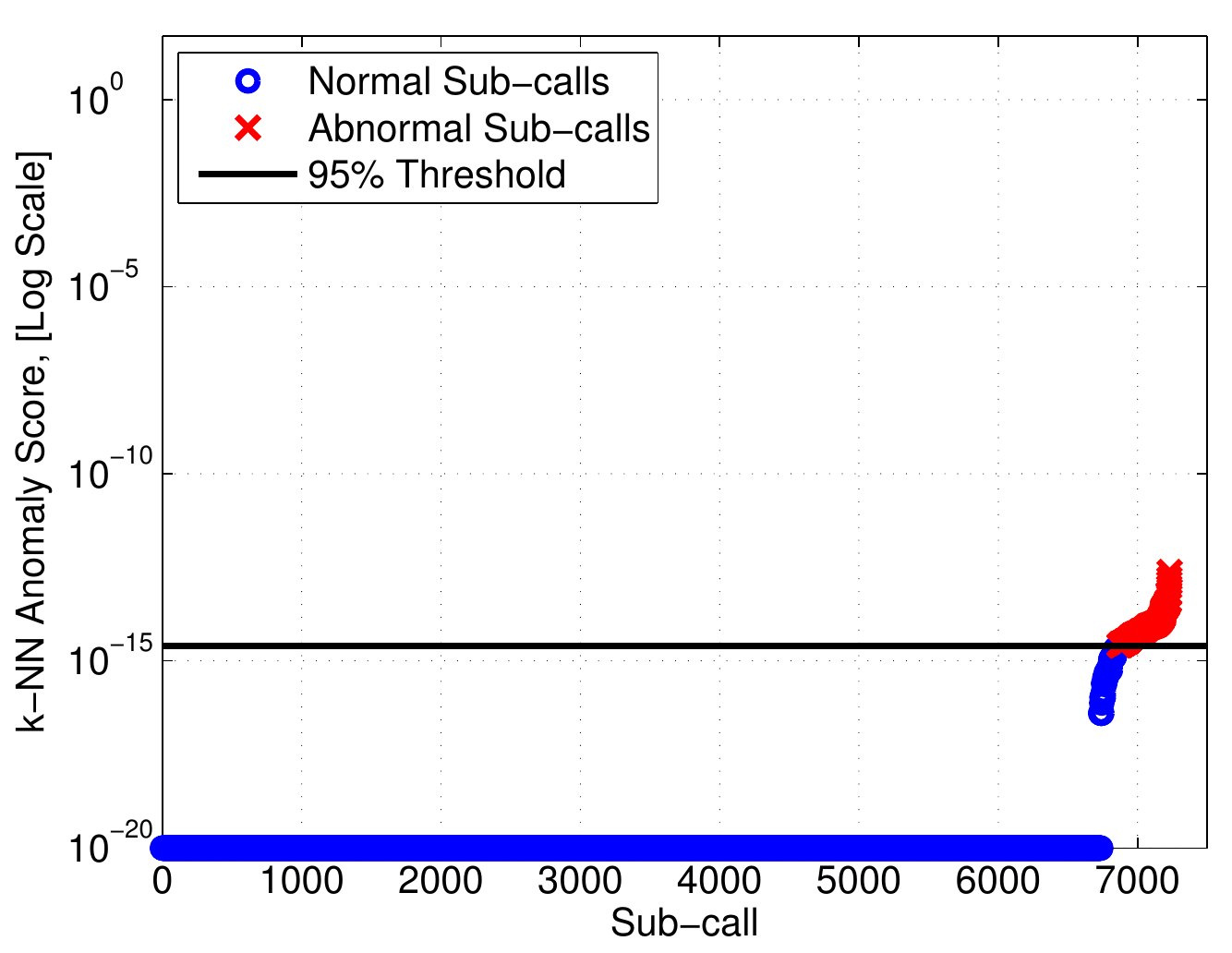}}
\caption{Normal dataset used for training of the sleeping cell detection framework}
\label{fig:normal_training_data}
\end{figure}

\begin{figure}
\centering
\subfloat[Problem testing dataset in the embedded space]{\label{fig:embedded_tst_prob}\includegraphics[scale=0.42]{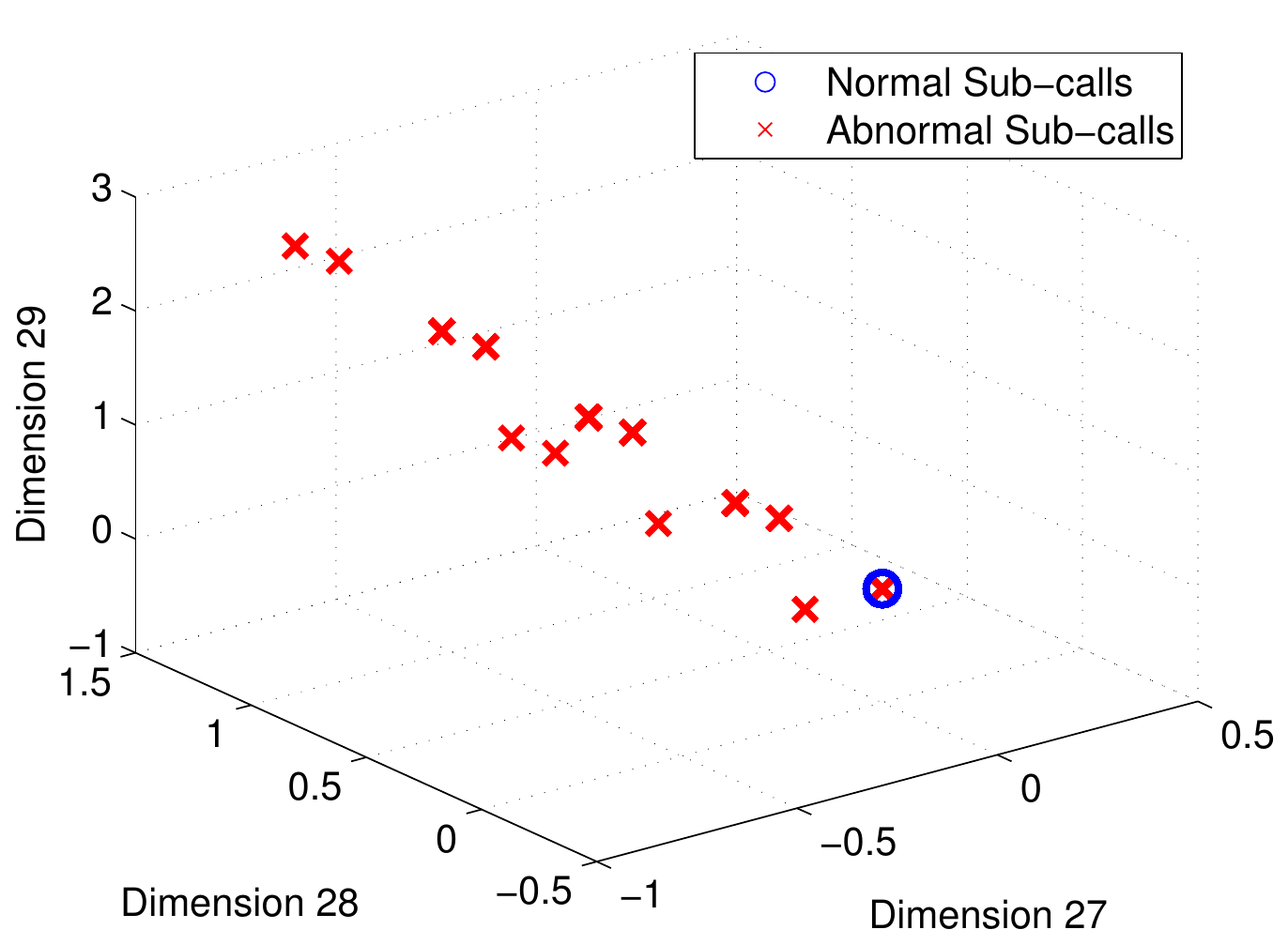}}
\subfloat[Sorted outlier scores of problem testing dataset]{\label{fig:knn_scores_prob}\includegraphics[scale=0.4]{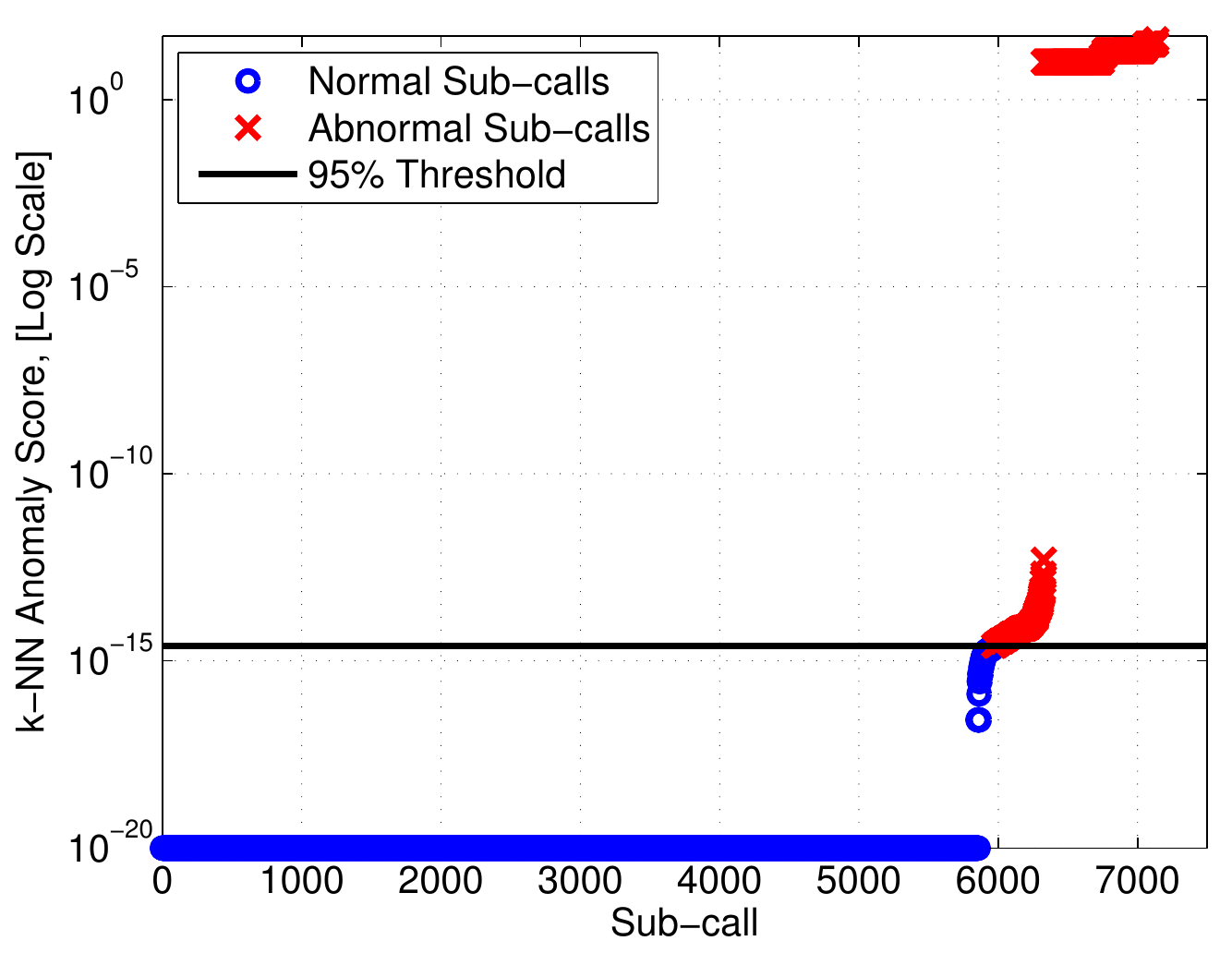}}
\caption{Problematic dataset used at the testing phase of the sleeping cell detection framework}
\label{fig:problematic_testing_data}
\end{figure}

After pre-processing with sliding window and N-gram methods, and transformation with \ac{mca}, training \ac{mdt} data is processed with \ac{knn} anomaly score algorithm. As it is discussed in section \ref{subsec:data_mining}, the anomaly score threshold, used for separation of data points to normal and abnormal classes, is selected to be $95^{th}$ percentile of outlier score in training data. 
Shape of normal training dataset in the embedded space is shown in Fig. 
\ref{fig:embedded_train_norm}, and sorted anomaly outlier 
scores are presented in Fig. \ref{fig:knn_scores_norm}. It can be seen that data points are very compact in the embedded space, and because of that there is no big difference in the anomaly score values. 
The main goals of analyzing testing dataset are to find anomalies, detect sleeping cell, and keep the false alarm rate as low as possible. 
At the testing phase either problematic or reference data are analyzed. After the same pre-processing stages as for training, the testing data is represented in the embedded space. When testing data is problematic dataset some of the samples are significantly further away from the main dense group of points, Fig.  \ref{fig:problematic_testing_data}. These abnormal points  are labeled as outliers, and the corresponding anomaly scores for these samples are much higher, as it can seen from Fig. \ref{fig:knn_scores_prob}. On the other hand, 
some of the points with relatively low anomaly score are above the abnormality threshold. This means that there is still certain percentage of false alarms, i.e. some ``good'' points are treated as ``bad''. The extent of negative effect caused by false alarms is discussed further in Section \ref{subsec:res_perf_eval}. Though, there is no 
opposite behavior referred to as ``miss-detection'' - none of the 
anomalous points are treated as normal.

\begin{figure}
\centering
\subfloat[Reference testing dataset in the embedded space]{\label{fig:embedded_tst_ref}\includegraphics[scale=0.42]{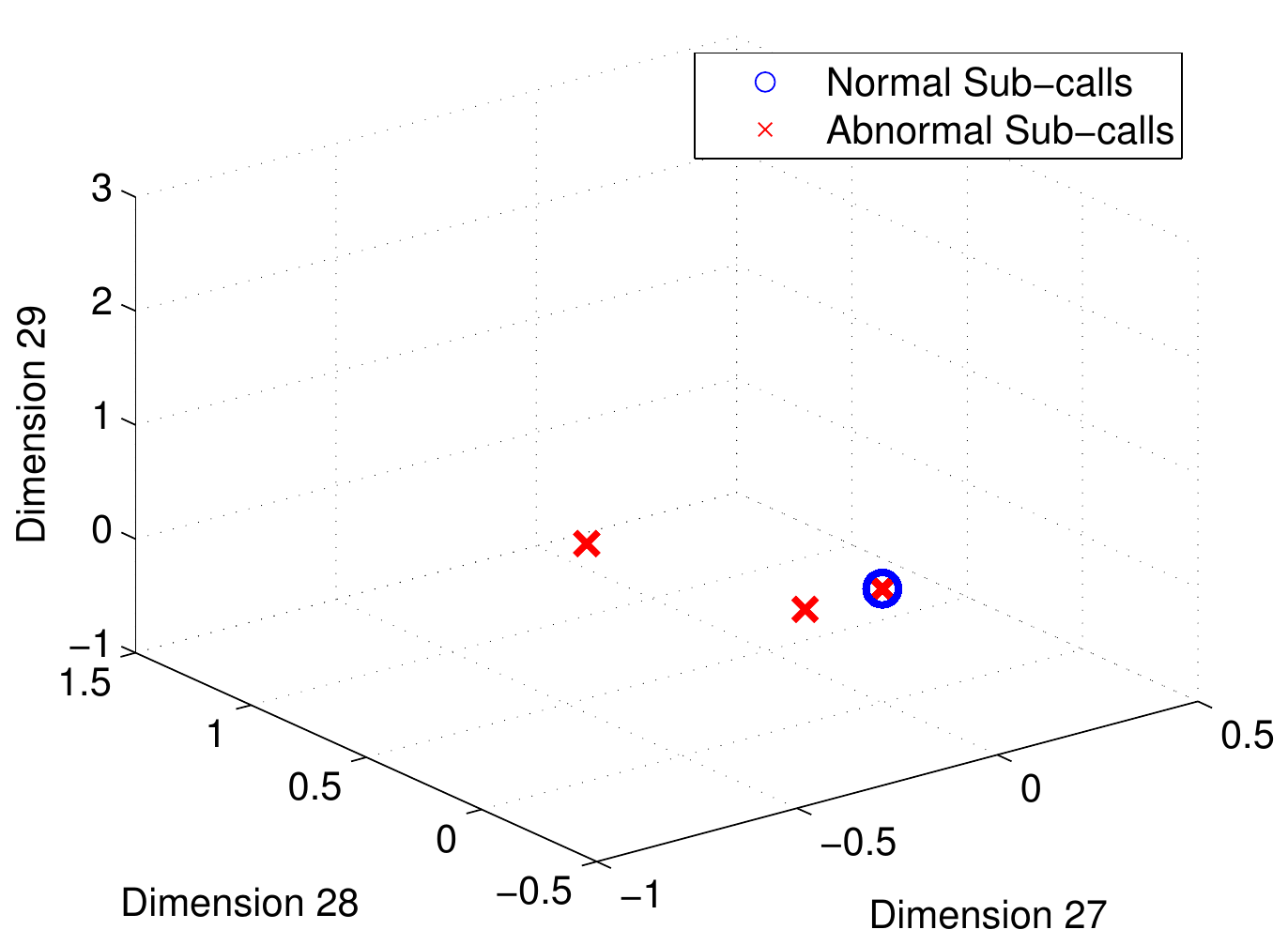}}
\subfloat[Sorted outlier scores of reference testing dataset]{\label{fig:knn_scores_ref}\includegraphics[scale=0.4]{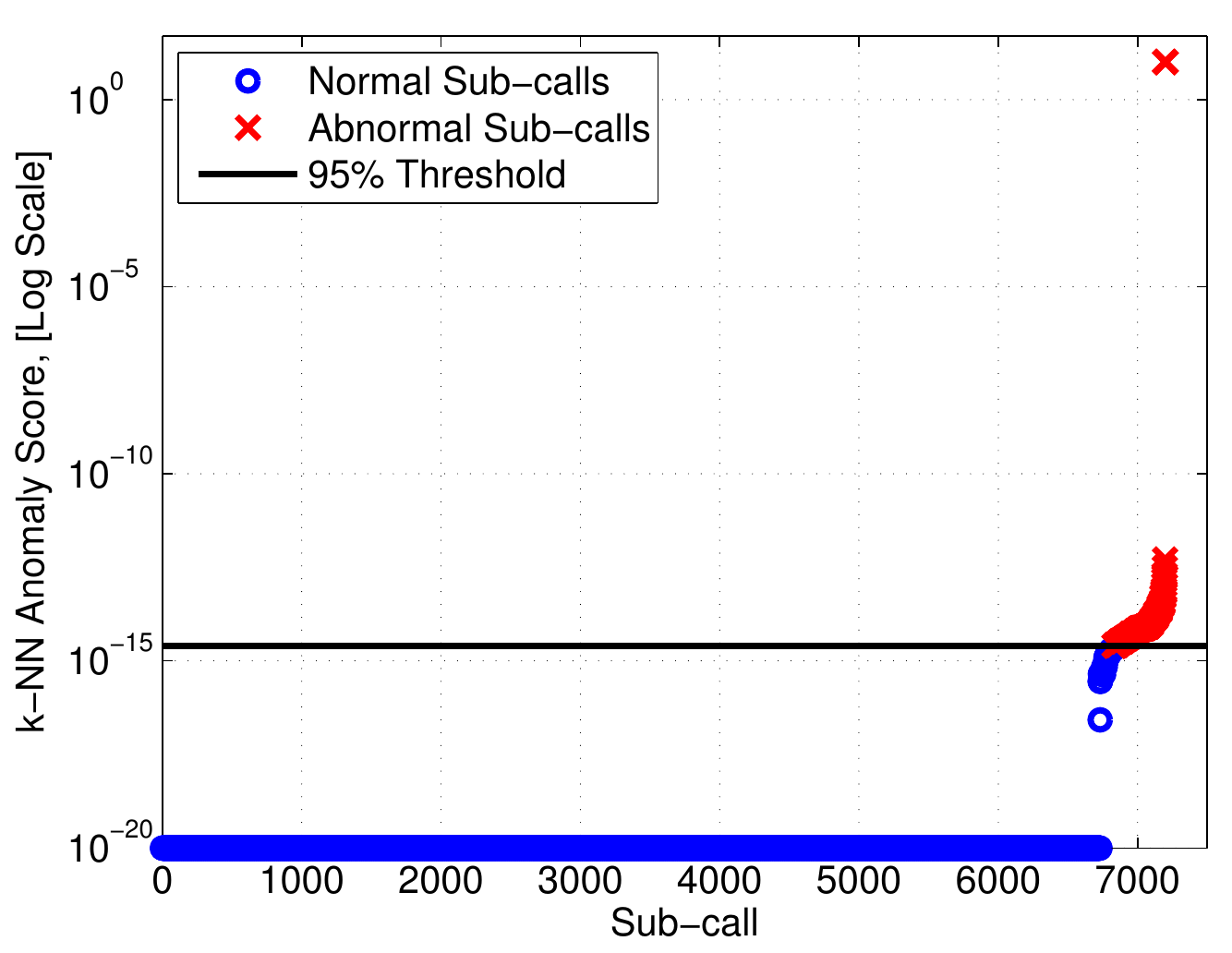}}
\caption{Reference dataset used at the testing phase of the sleeping cell detection framework}
\label{fig:reference_testing_data}
\end{figure}

Validation of the data mining framework is done by using error-free reference dataset as testing data. No real anomalies are present in the network behavior. Reference testing data in the embedded space and corresponding anomaly outlier scores are shown in Fig. \ref{fig:reference_testing_data}. Only few points can be treated as outliers, and in general the shapes of normal (Fig. \ref{fig:embedded_train_norm}) and 
reference (Fig. \ref{fig:embedded_tst_ref}) datasets in the embedded 
space are very similar. 
Anomaly outlier scores of the reference testing data is low for all points, except 2 outliers.  

\subsection{Application of Post-Processing Methods for Sleeping Cell Detection}
\label{subsec:res_postproc_methods}
After training and testing phases certain sub-calls are marked as anomalies. The next step is conversion of this information to knowledge about location of malfunctioning cell or cells, and this is done through post-processing described in Section \ref{subsec:post_processing}. 

\subsubsection{Detection based on Dominance Cell Sub-Call Deviation}
\label{subsub:res_visits_dev}
\begin{figure}
\centering
\subfloat[Problematic dataset sleeping cell detection histogram]{\label{fig:visits_dev_hist_prob}\includegraphics[scale=0.47]{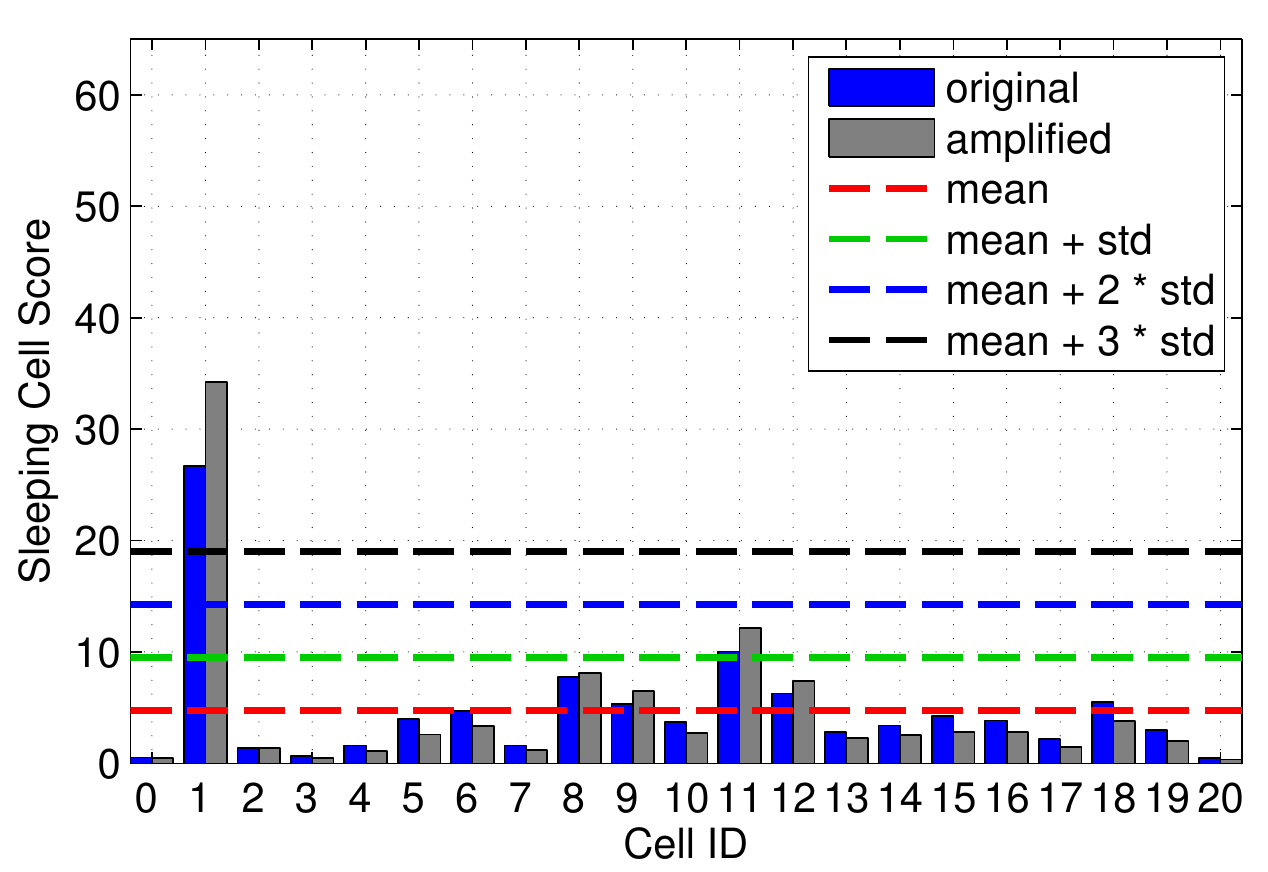}
}
\subfloat[Reference dataset sleeping cell detection histogram]{\label{fig:visits_dev_hist_ref}\includegraphics[scale=0.47]{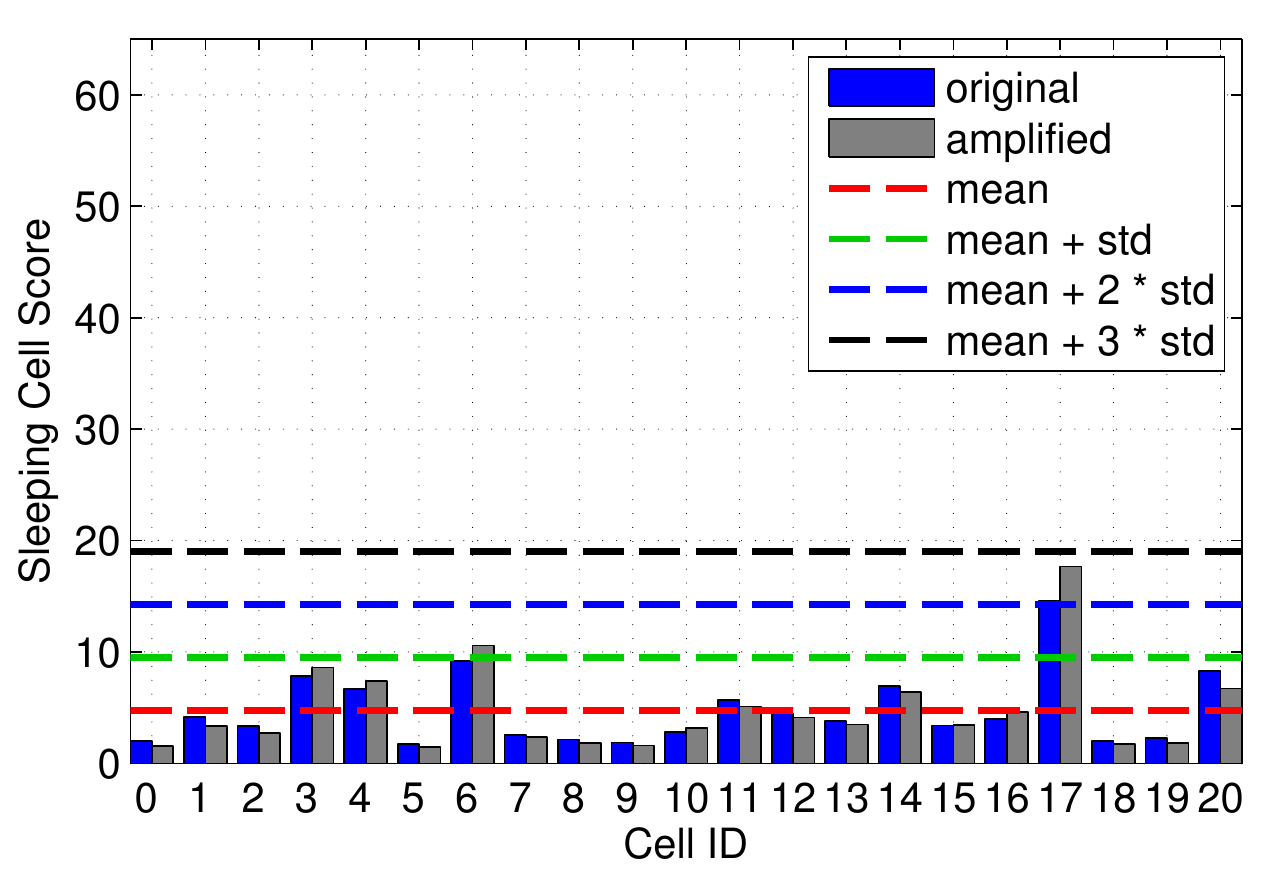}
} 
\\
\subfloat[Problematic dataset heat map]{\label{fig:visits_dev_heat_map}\includegraphics[scale=0.42]{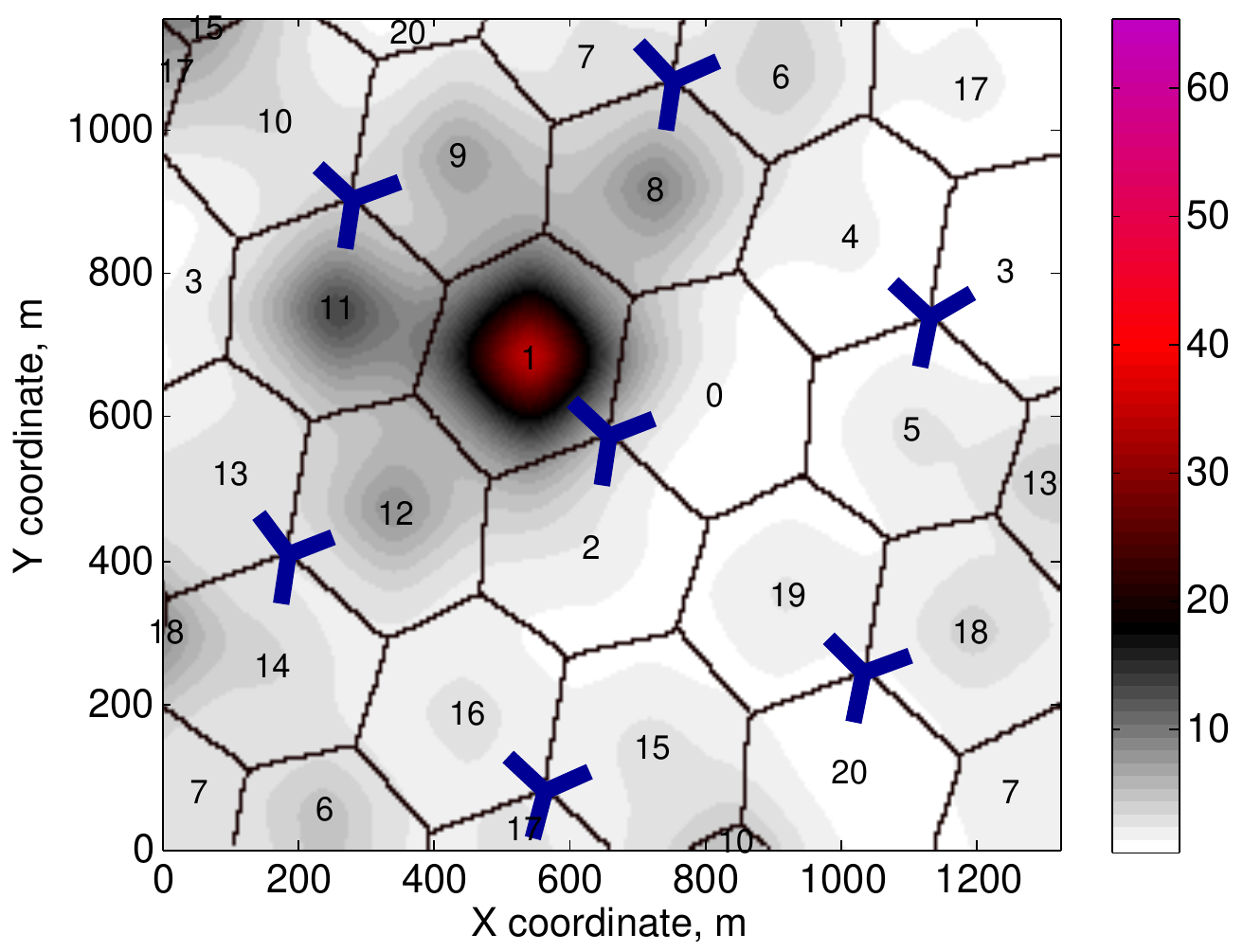}
} 
\subfloat[Reference dataset heat map]{\label{fig:visits_dev_hist_ref_heat_map}\includegraphics[scale=0.42]{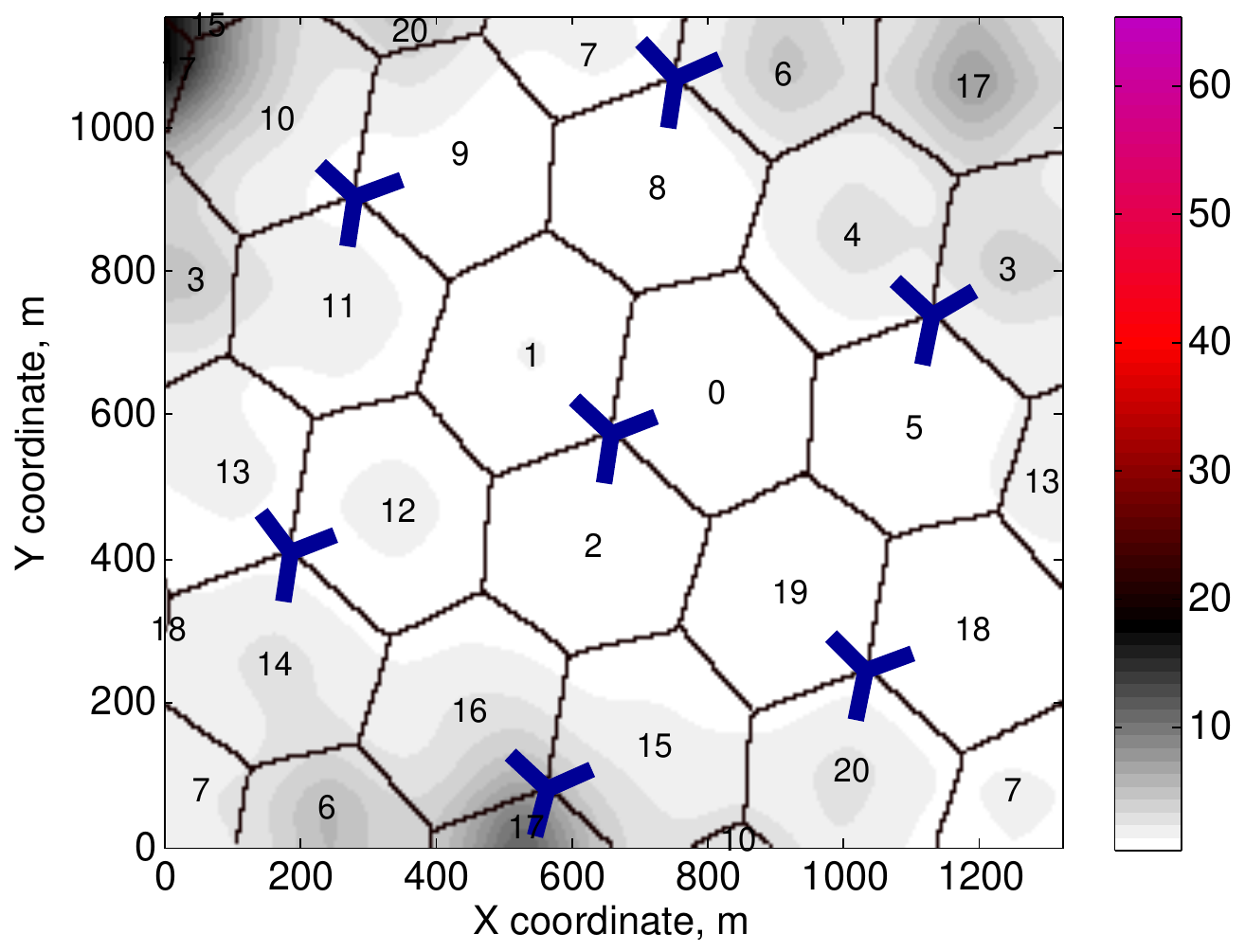}
}
\caption{Results of sleeping cell detection for Dominance Cell Sub-Call Deviation method}
\label{fig:res_visits_dev_prob}
\end{figure}

In our earlier study \cite{Chernogorov13} post-processing based on dominance cells and call deviation for 
sleeping cell detection is presented. One problem of using calls as samples is that, in case if the duration of the analyzed user call is long, the 
corresponding number of visited cells is large, especially for fast 
\acp{ue}. Hence, even if certain call is classified as abnormal, it is very hard to 
say which cell has anomalous behavior. To 
overcome this problem, analysis is done for sub-calls, derived with sliding window method, see Section \ref{subsub:sliding_window}. Majority of sub-calls contain the same number of network events, and the length of the analyzed sequence is short enough to identify the exact cell, with problematic behavior.
Deviation measures the difference between training and testing data, and it is used to sleeping cell 
detection histogram, presented in Fig. \ref{fig:visits_dev_hist_prob}. 
From this figure, it can be seen that abnormal sub-calls are encountered more frequently 
in the area of dominance of cell 1, which has the highest deviation. One can see that there are 2 types of bars - 
colorful (in this case blue) and grey. The second variant implies 
additional post-processing step - amplification, described in 
Section \ref{subsec:post_processing}. In addition to cell 1, its neighboring cells 8, 9, 11 and 12 also have 
increased deviation values, as it can be seen from the network heat map in Fig. \ref{fig:visits_dev_heat_map}.
Sleeping cell detection histogram and network heat map for reference dataset used as testing are shown in Fig. \ref{fig:visits_dev_hist_ref} and \ref{fig:visits_dev_hist_ref_heat_map} correspondingly. Even though cells 6 and 17 have higher \ac{sc} scores than other cells, they are not marked as abnormal, because their abnormality does not reach mean + $3\sigma$ level.
\subsubsection{Detection based on Dominance Cell 2-Gram Deviation}
\label{subsub:res_occurence_dev}
\begin{figure}
\centering
\subfloat[Problematic dataset sleeping cell detection histogram]{\label{fig:occurence_dev_hist_prob}\includegraphics[scale=0.47]{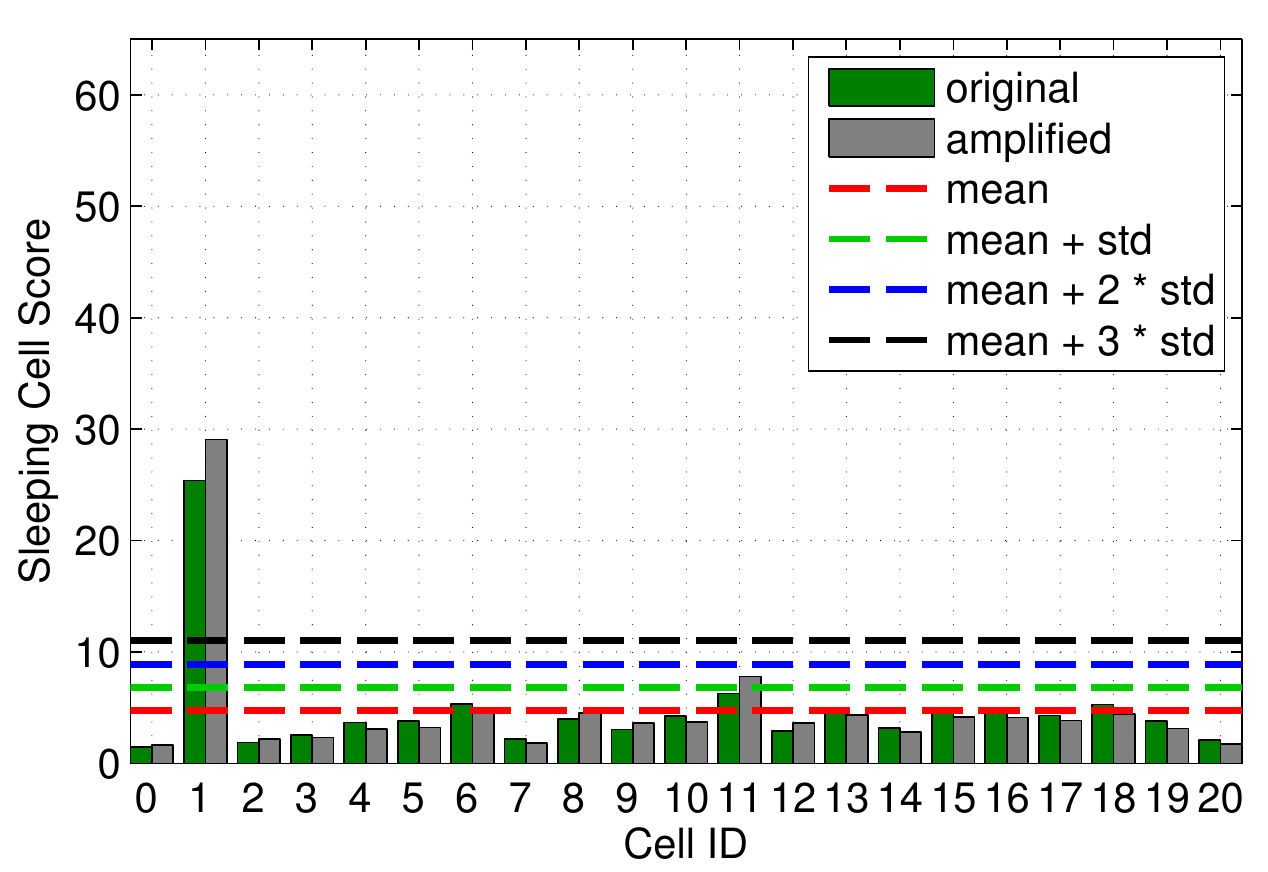}
}
\subfloat[Reference dataset sleeping cell detection histogram]{\label{fig:occurence_dev_hist_ref}\includegraphics[scale=0.47]{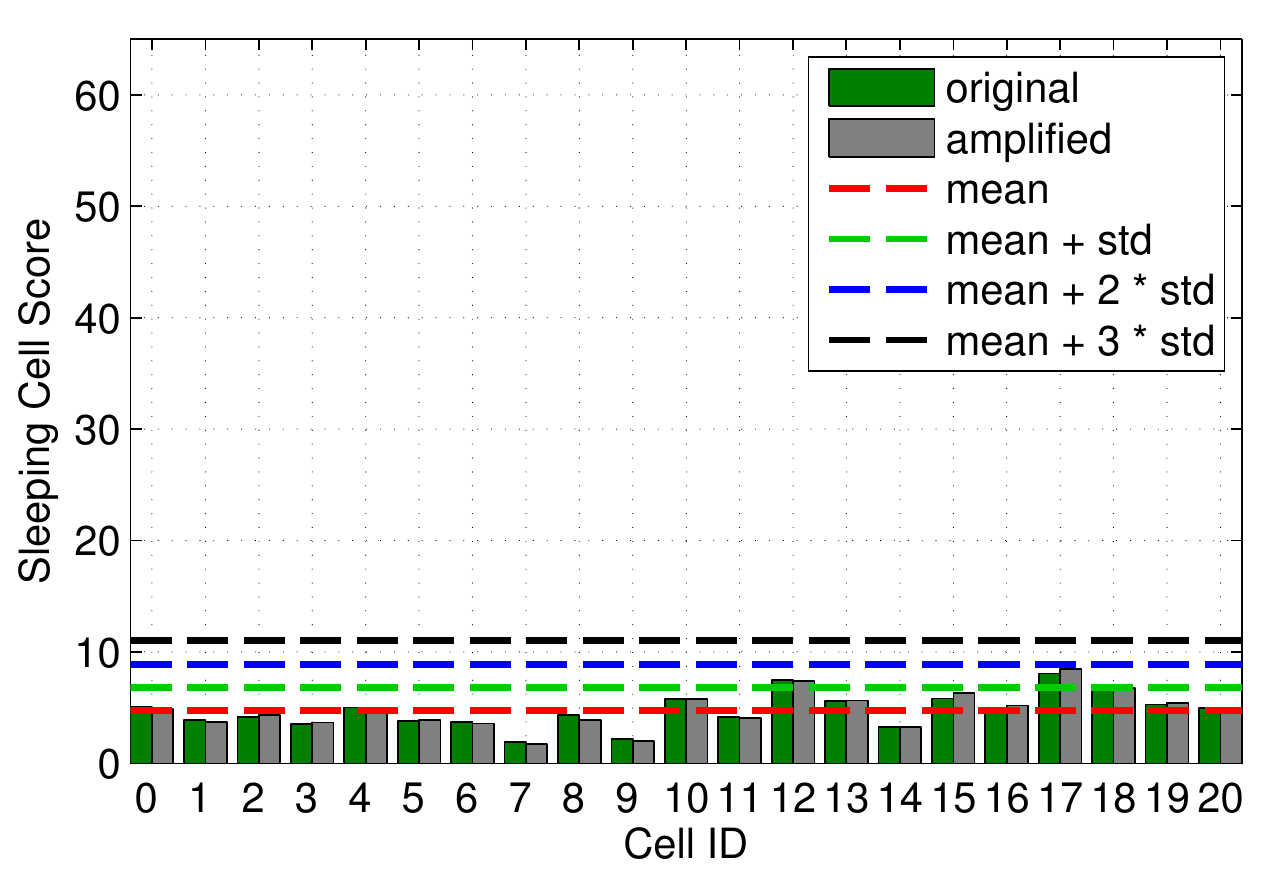}
}
\\
\subfloat[Problematic dataset heat map]{\label{fig:occurence_dev_heat_map_prob}\includegraphics[scale=0.42]{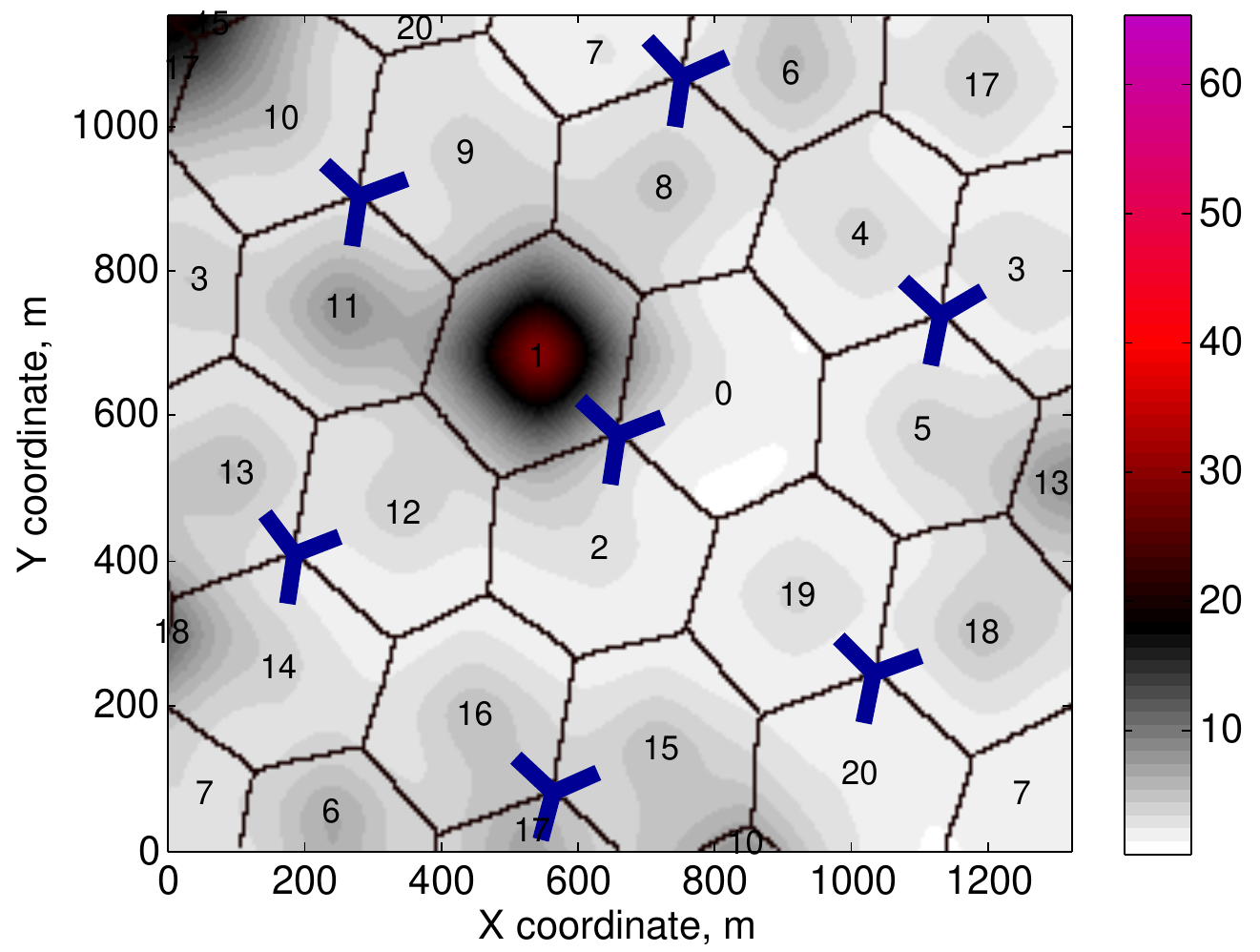}
}
\subfloat[Reference dataset heat map]{\label{fig:occurence_dev_heat_map_ref}\includegraphics[scale=0.42]{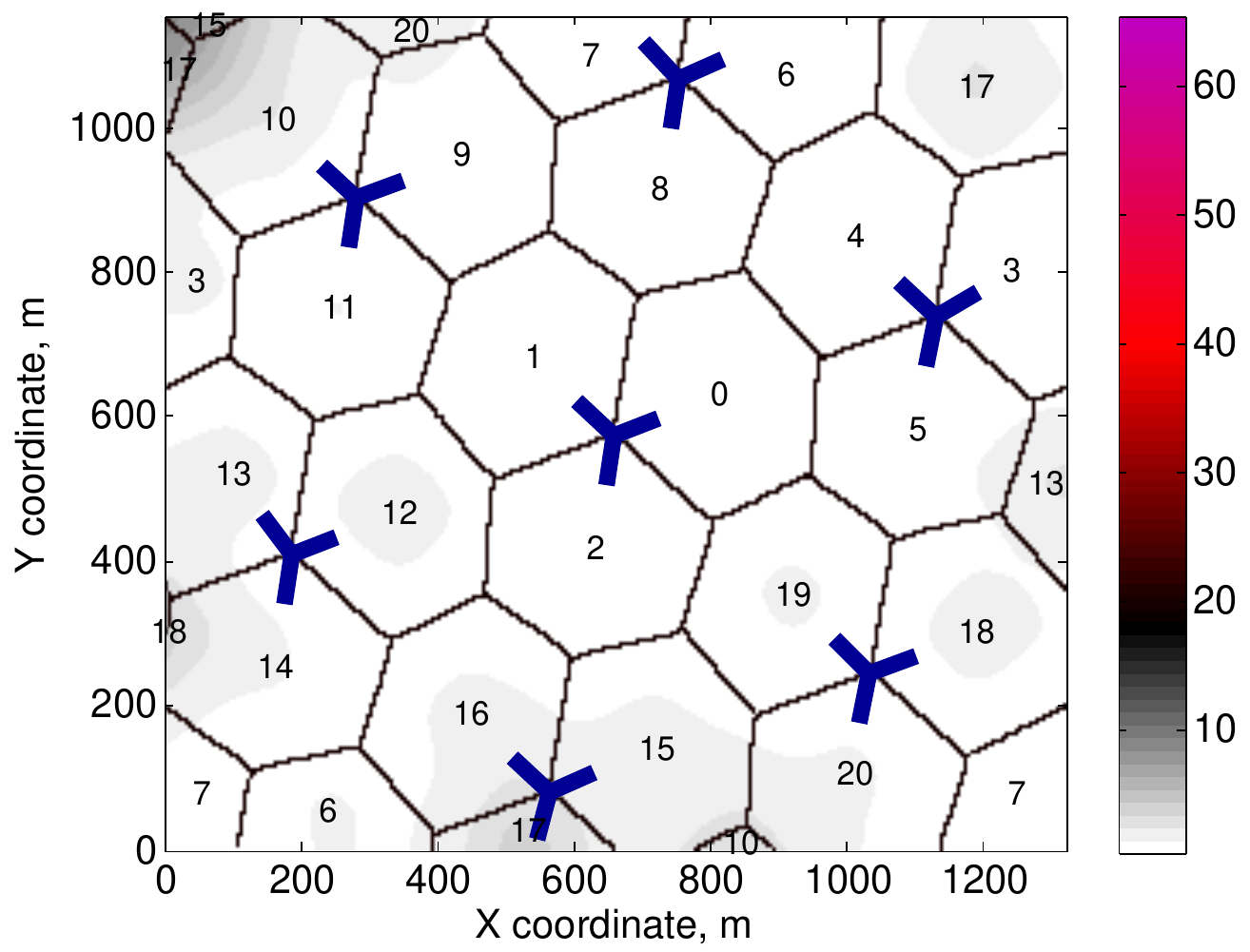}
}
\caption{Results of sleeping cell detection for Dominance Cell 2-Gram Deviation method}
\label{fig:res_occurence_dev_prob}
\end{figure}

In this method problematic network regions are found through comparison of occurrence frequencies, normalized by the total number of users, in training and testing datasets. In case there is a big increase or decrease, the cell associated with these changes is marked as abnormal. 
From sleeping cell detection histogram in Fig. 
\ref{fig:occurence_dev_hist_prob} it can be that cell 1 has a clear 
difference in number of 2-gram occurrences in testing data, if compared to training 
data. This happens because handovers toward this cell fail. Due to 
this fact 2-gram sequence with events related to handovers become 
imbalanced in testing data if compared to training data. For instance, 
2-grams like \ac{ho} Command - \ac{ho} Complete and 
\ac{ho} Complete - A2 RSRP ENTER, become very rare. On the other hand, 2-
gram \ac{ho} Command - A2 RSRP ENTER, which can be treated as indication of non-successful handovers, in opposite becomes very popular in testing 
data, while in training data it does not exist at all. Among the 
neighbors of problematic cell 1, only cell 11 has slightly increased 
sleeping cell score.
Testing sleeping cell detection framework with reference data and 
post-processing with Dominance Cell 2-Gram 
Deviation method demonstrates lower false-alarm rate than Dominance Cell 
Sub-Call Deviation, as it can be seen from Fig. 
\ref{fig:occurence_dev_hist_ref} and \ref{fig:occurence_dev_heat_map_ref}. 

\subsubsection{Detection based on Dominance Cell 2-Gram Symmetry Deviation}
\label{subsub:res_symmetry_dev}
This post-processing method analyzes the symmetry imbalance of network events 2-grams. Information about number of 2-gram directed to the cell, and from the cell is extracted from the training. In case if in the testing data the balance (number of 2-grams, which start in this cell, )Thus, only 2-grams, which occur at cell borders, i.e. in the dominance area of 2 cells, are considered. It means that if in the training data, the number of handovers from Cell A to Cell B, and from Cell B to Cell A, is roughly the same, this cell has balanced 2-gram 
it can concluded that symmetry of this particular 2-gram is skewed. disturbed 
comparing to the training data. Most common types of 2-grams which are 
analyzed with this method are related to handovers, e.g. A3 - \ac{ho} 
COMMAND sequences.

\begin{figure}
\centering
\subfloat[Problematic dataset sleeping cell detection histogram]{\label{fig:symmetry_hist_prob}\includegraphics[scale=0.47]{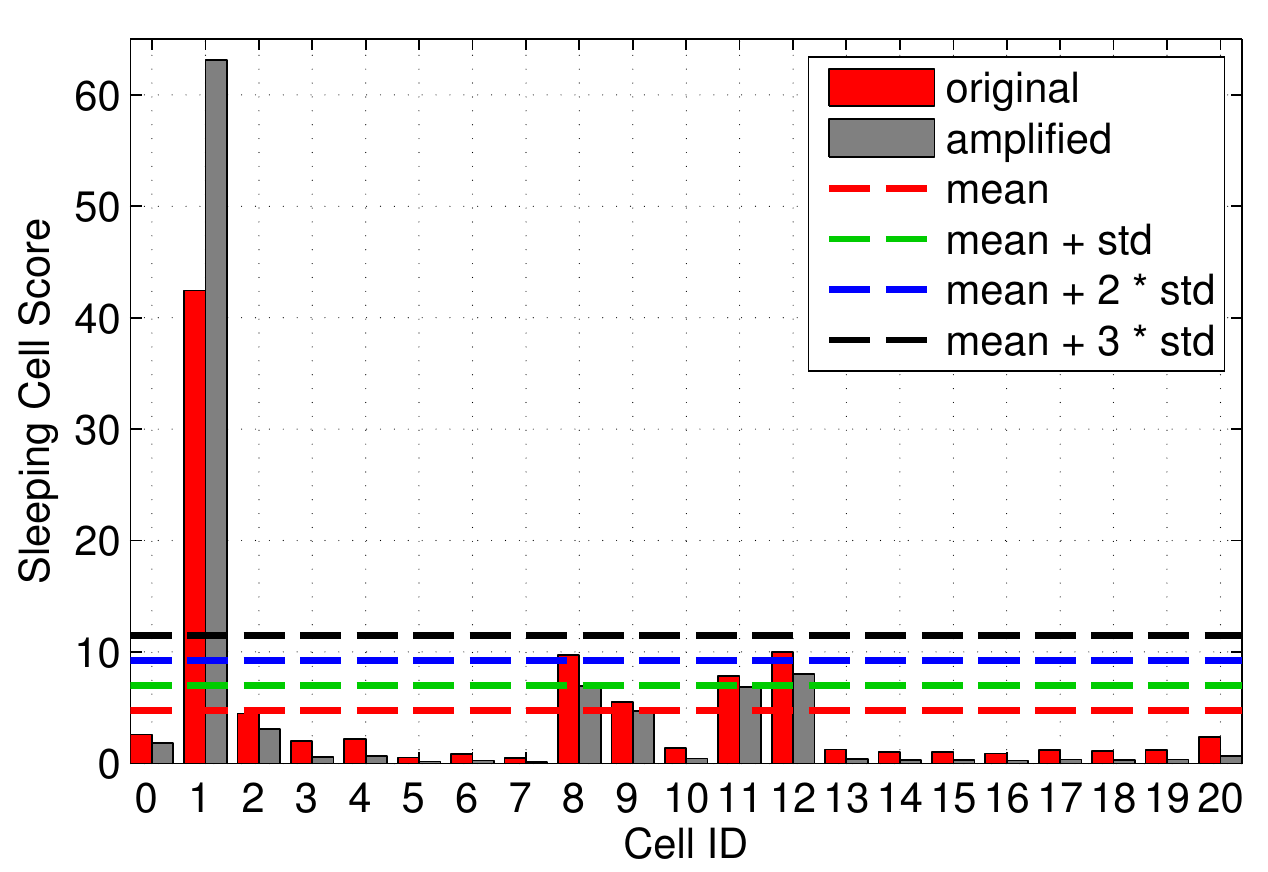}
}
\subfloat[Reference dataset sleeping cell detection histogram]{\label{fig:symmetry_hist_ref}\includegraphics[scale=0.47]{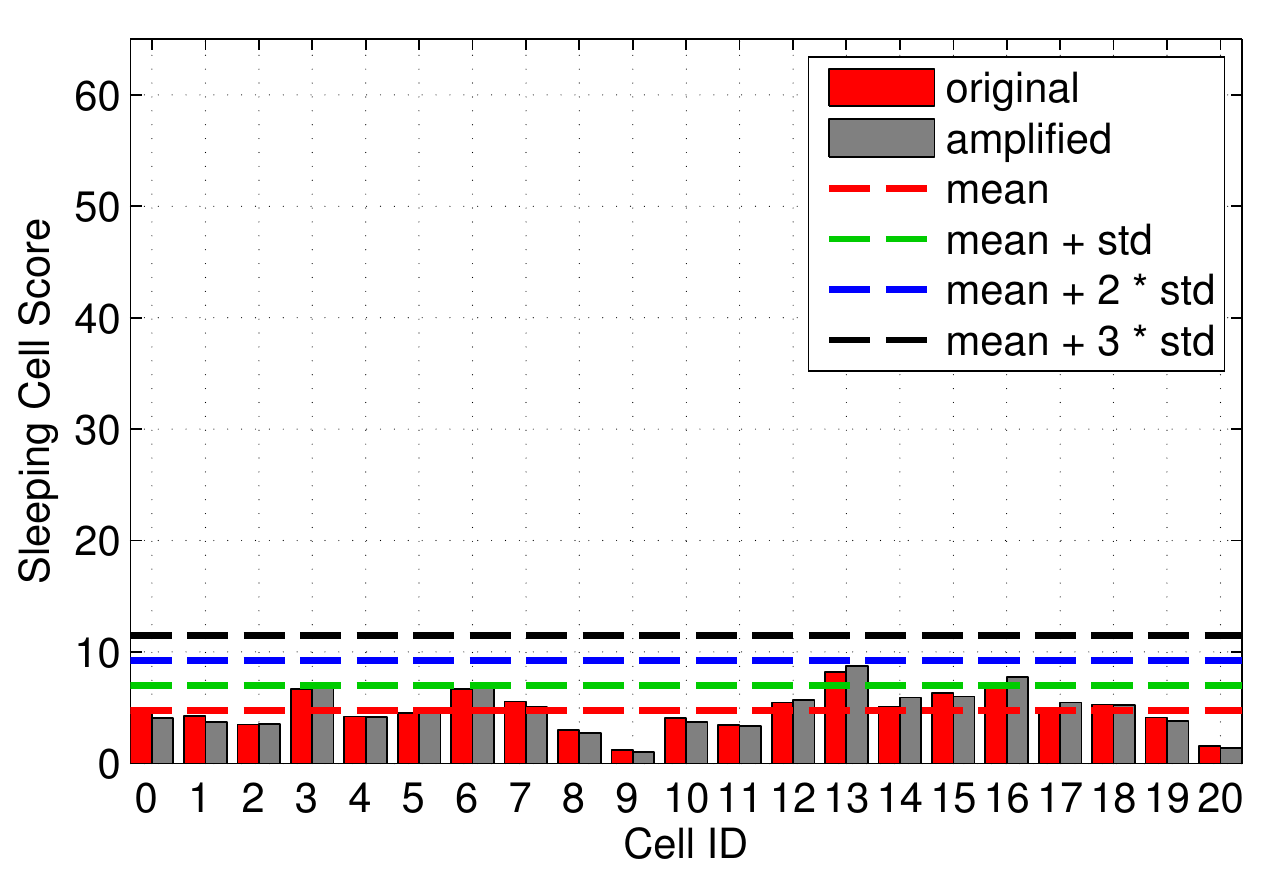}
}
\\
\subfloat[Problematic dataset heat map]{\label{fig:symmetry_heat_map_prob}\includegraphics[scale=0.42]{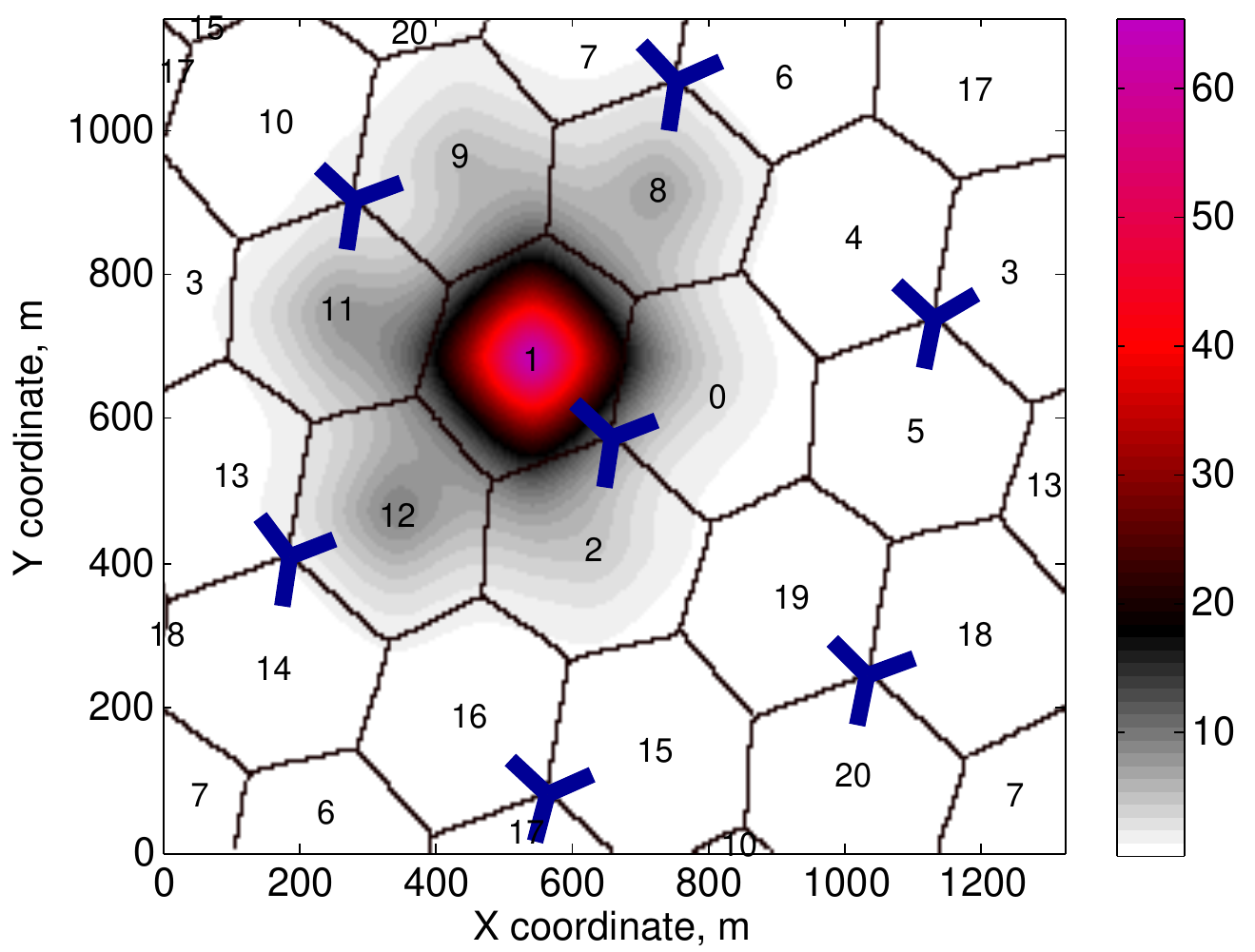}
}
\subfloat[Reference dataset heat map]{\label{fig:symmetry_heat_map_ref}\includegraphics[scale=0.42]{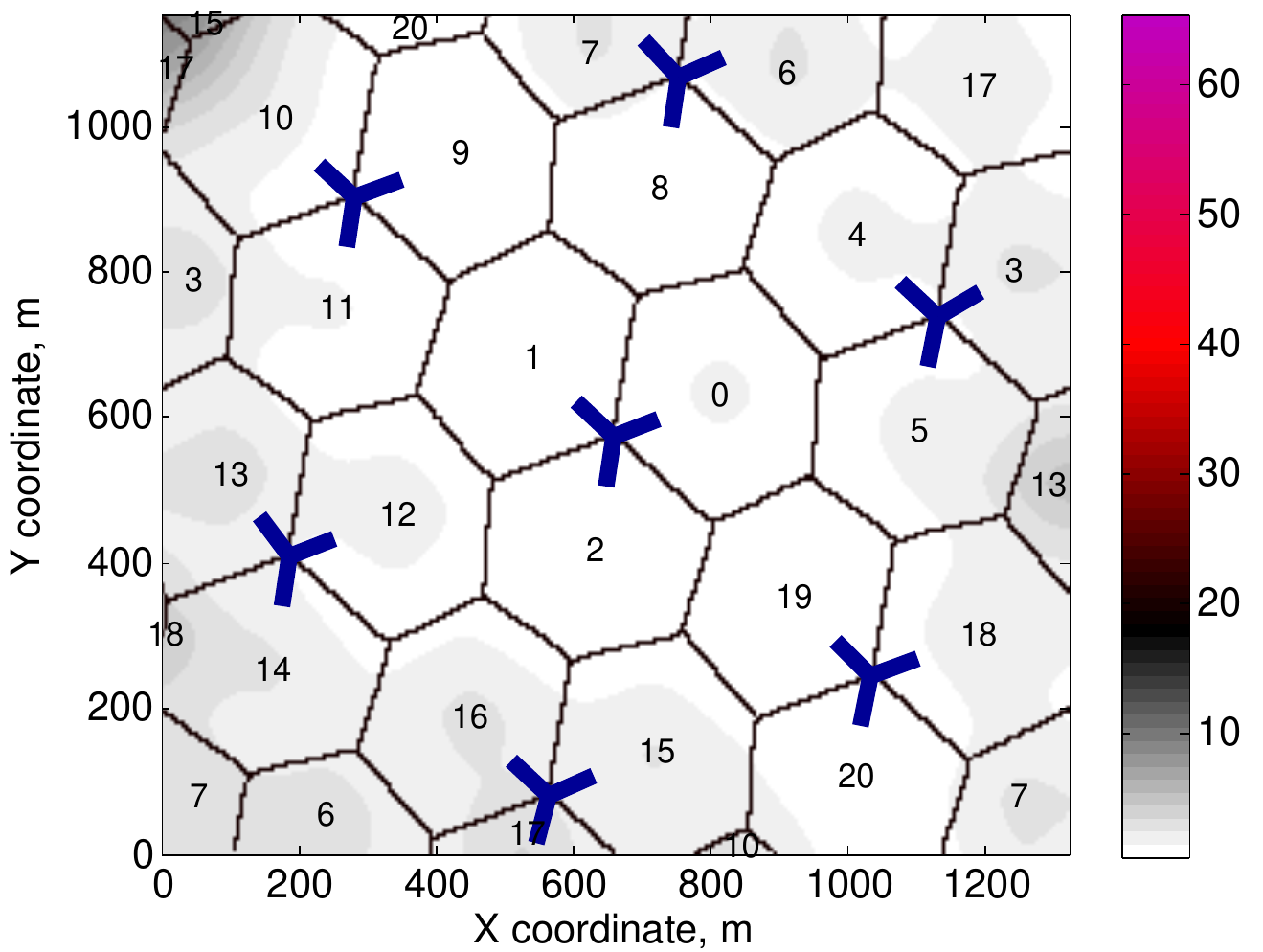}
}
\caption{Results of sleeping cell detection for Dominance Cell 2-Gram Symmetry Deviation method}
\label{fig:res_symmetry_prob}
\end{figure}

From Fig. \ref{fig:res_symmetry_prob} it can be seen that Dominance 
Cell 2-Gram Symmetry Deviation finds sleeping cell 1, while its 
neighboring cells 8, 9, 11 and 12 have suspiciously high sleeping cell 
score, if compared to other cells in the network.

Comparison of symmetry analysis method with two previously described
post-processing approaches shows that this method is very efficient in 
detecting sleeping cell and its neighbors. At the same time stability, i.e. false alarm rate, of this method is also very good, as it can be seen from Fig. \ref{fig:symmetry_hist_ref}.

\subsubsection{Detection based on Target Cell Sub-Calls}
\label{subsub:res_trg_cell_dev}
\begin{figure}
\centering
\subfloat[Problematic dataset sleeping cell detection histogram]{\label{fig:trg_cell_hist_prob}\includegraphics[scale=0.47]{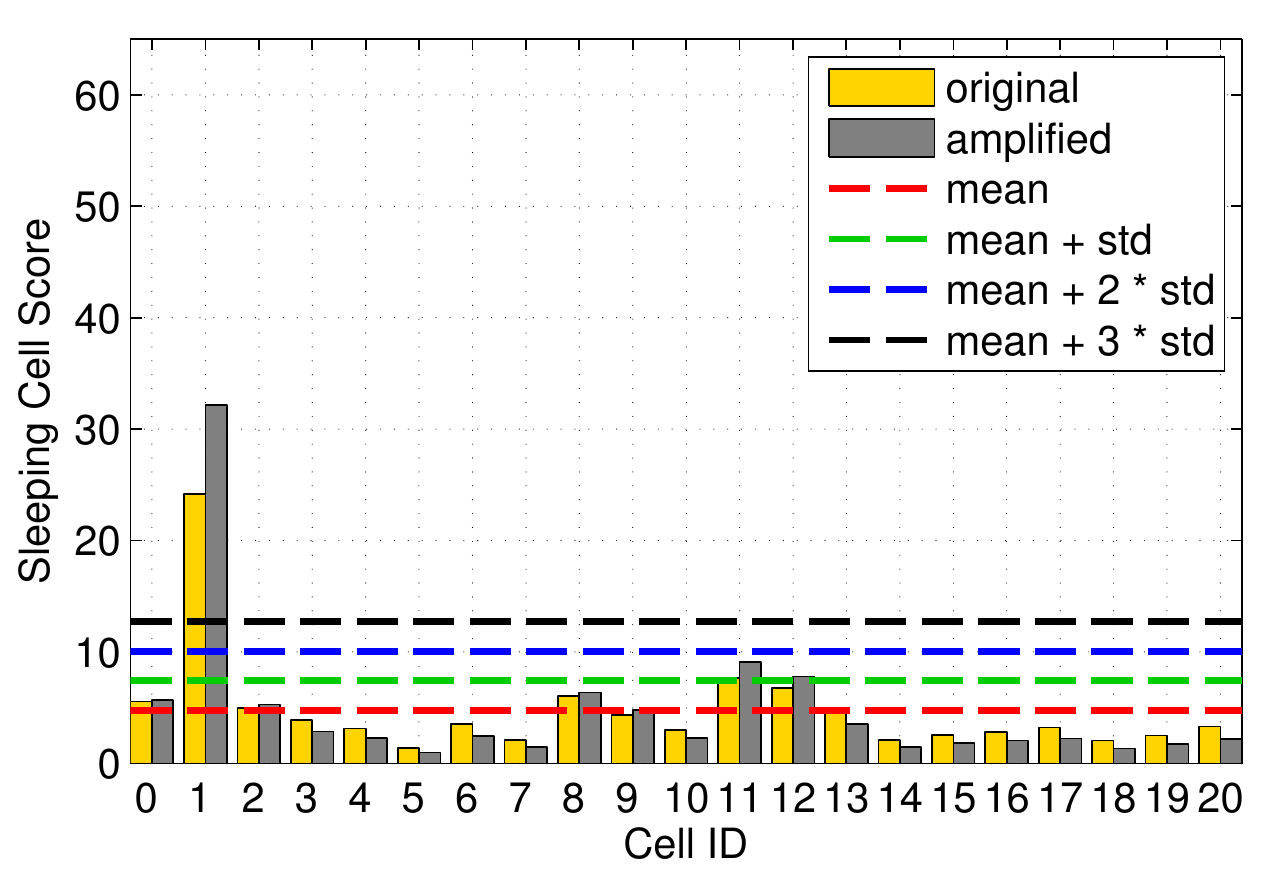}
}
\subfloat[Reference dataset sleeping cell detection histogram]{\label{fig:trg_cell_hist_ref}\includegraphics[scale=0.47]{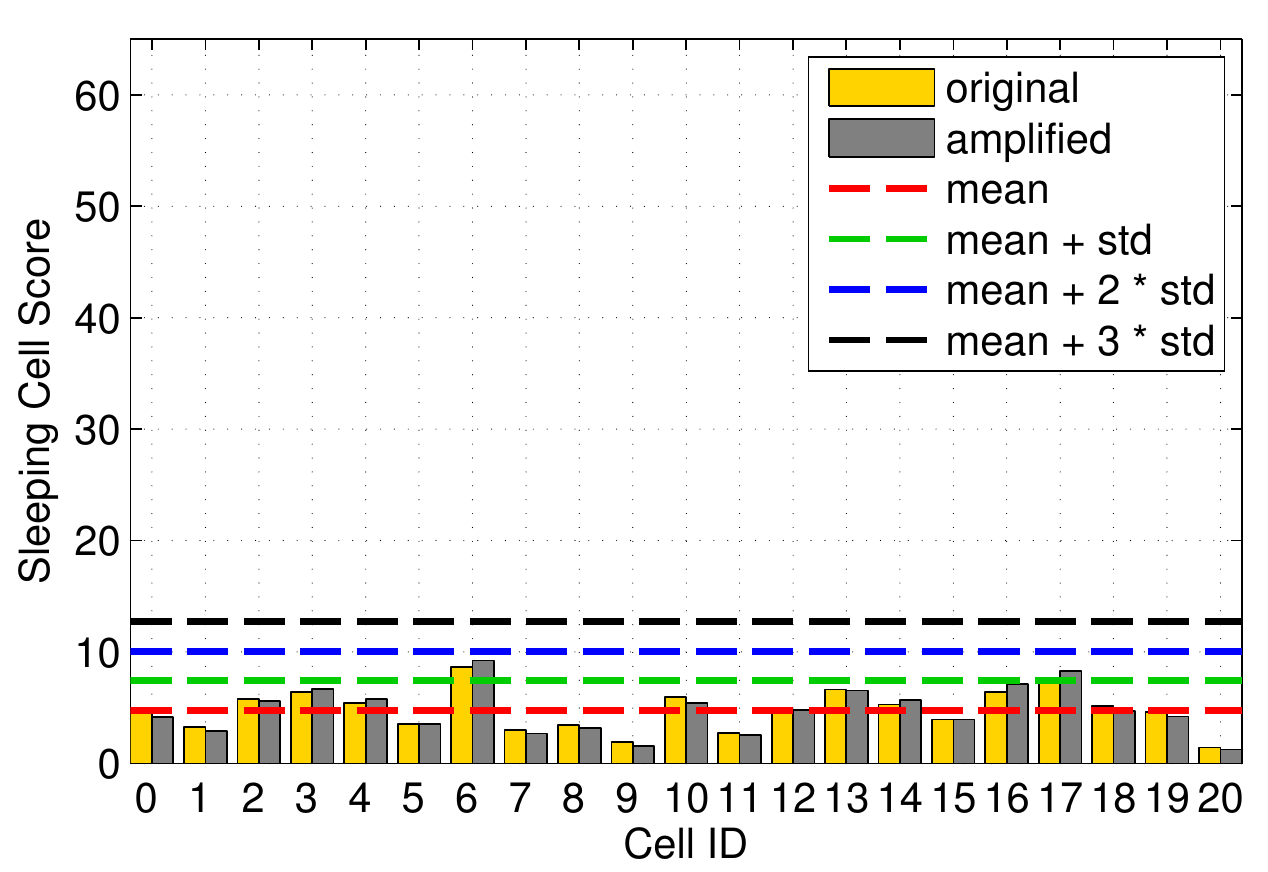}
}
\\
\subfloat[Problematic dataset heat map]{\label{fig:trg_cell_heat_map_prob}\includegraphics[scale=0.42]{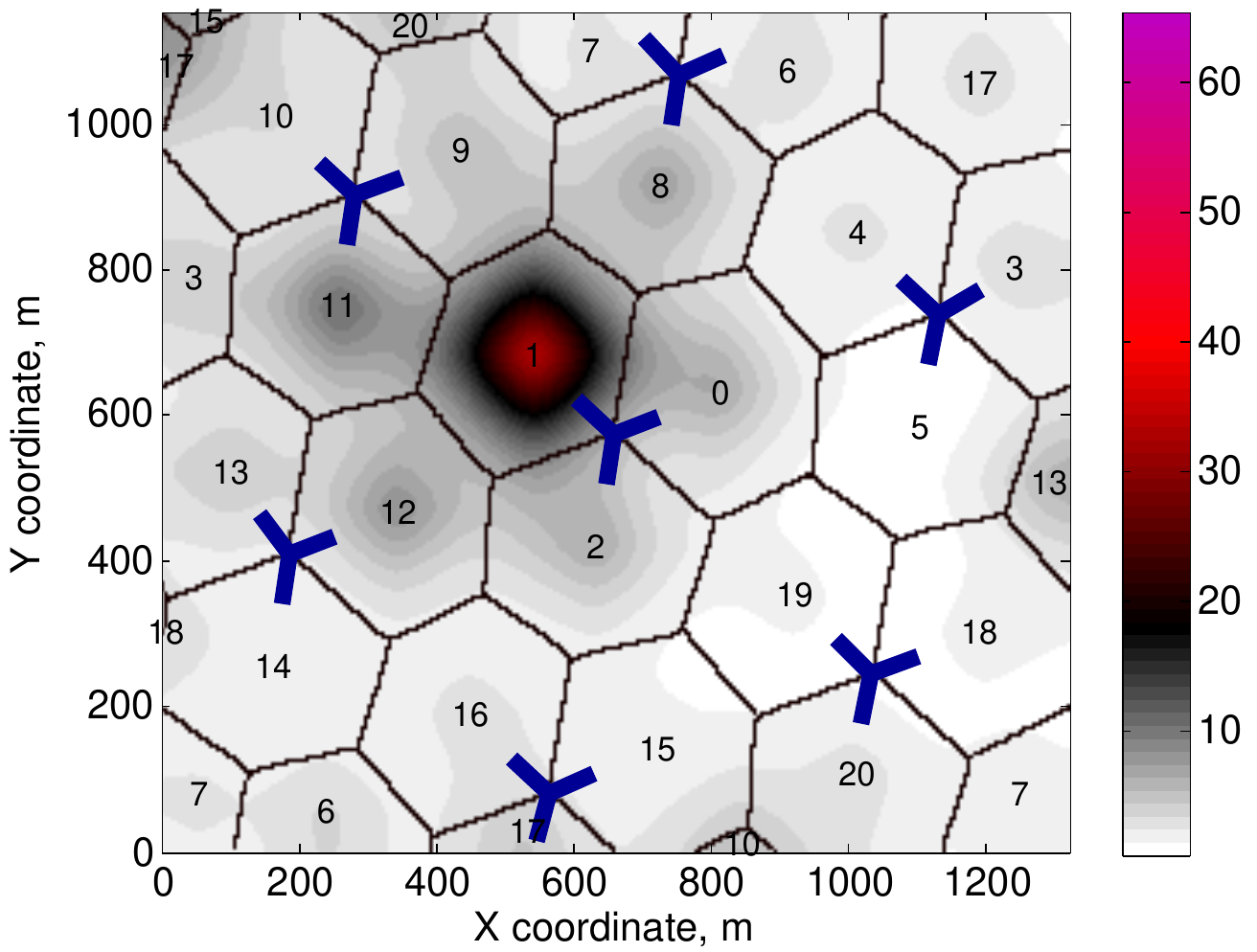}
}
\subfloat[Reference dataset heat map]{\label{fig:trg_cell_heat_map_ref}\includegraphics[scale=0.42]{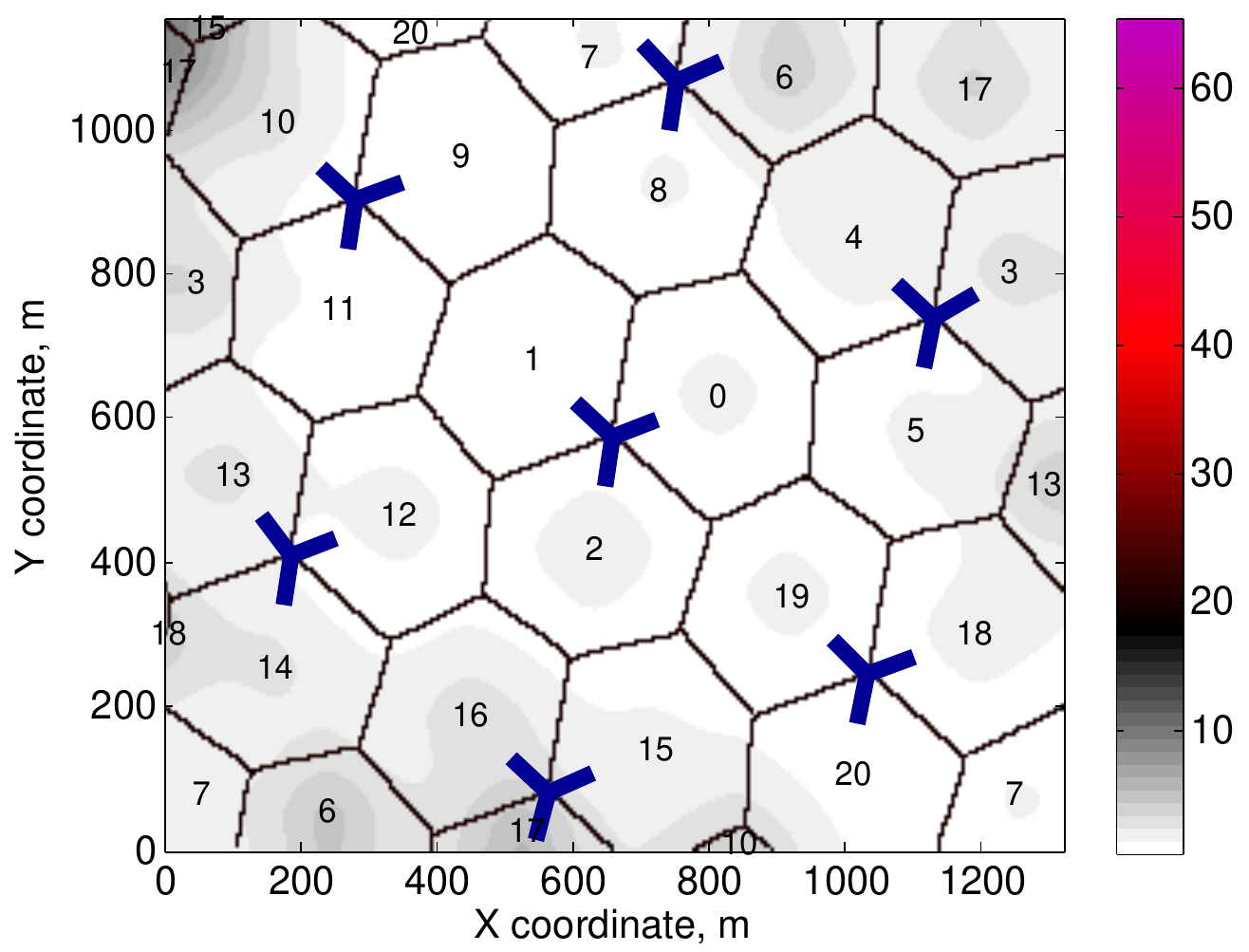}
}
\caption{Results of sleeping cell detection for Target Cell Sub-Calls method}
\label{fig:res_trg_cell_prob}
\end{figure}

As it is discussed in Section \ref{subsec:post_processing}, deviation between training and testing data is 
not calculated in this method. Extensive location information, like dominance map information, is not required for sleeping cell detection  with target cell sub-call method.
The sleeping cell detection histogram, presented in Fig. 
\ref{fig:res_trg_cell_prob}, is constructed by counting all unique target cell IDs for each anomalous sub-call. It can be clearly seen that cell 1 is successfully detected. Neighboring cells 8, 9, 11 and 12 also contain indication of malfunction in this area, as it can be noticed from heat map, shown in Fig. \ref{fig:trg_cell_hist_ref}.
For this method, the \ac{sc} score of cell 1 is slightly lower than for the post-processing methods, based on dominance cell deviation. Another shortcoming is that target cell sub-call method is more prone to trigger false alarms. This can be seen from the results when reference data is used as testing, Fig. 
\ref{fig:trg_cell_hist_ref}. Sleeping cell score of cell 6 is 
reaching threshold of mean plus 2 standard deviations. For cells 16 and 17 \ac{sc} scores are also quite high, as it can also be noticed from Fig. \ref{fig:trg_cell_heat_map_ref}. On the other hand, target cell sub-call method is much simpler, and requires significantly less information about user event occurrence location.

\subsection{Combined Method of Sleeping Cell Detection}
\label{subsec:res_combined}
\begin{figure}
\centering
\subfloat[Problematic dataset sleeping cell detection histogram]{\label{fig:combined_hist_prob}\includegraphics[scale=0.44]{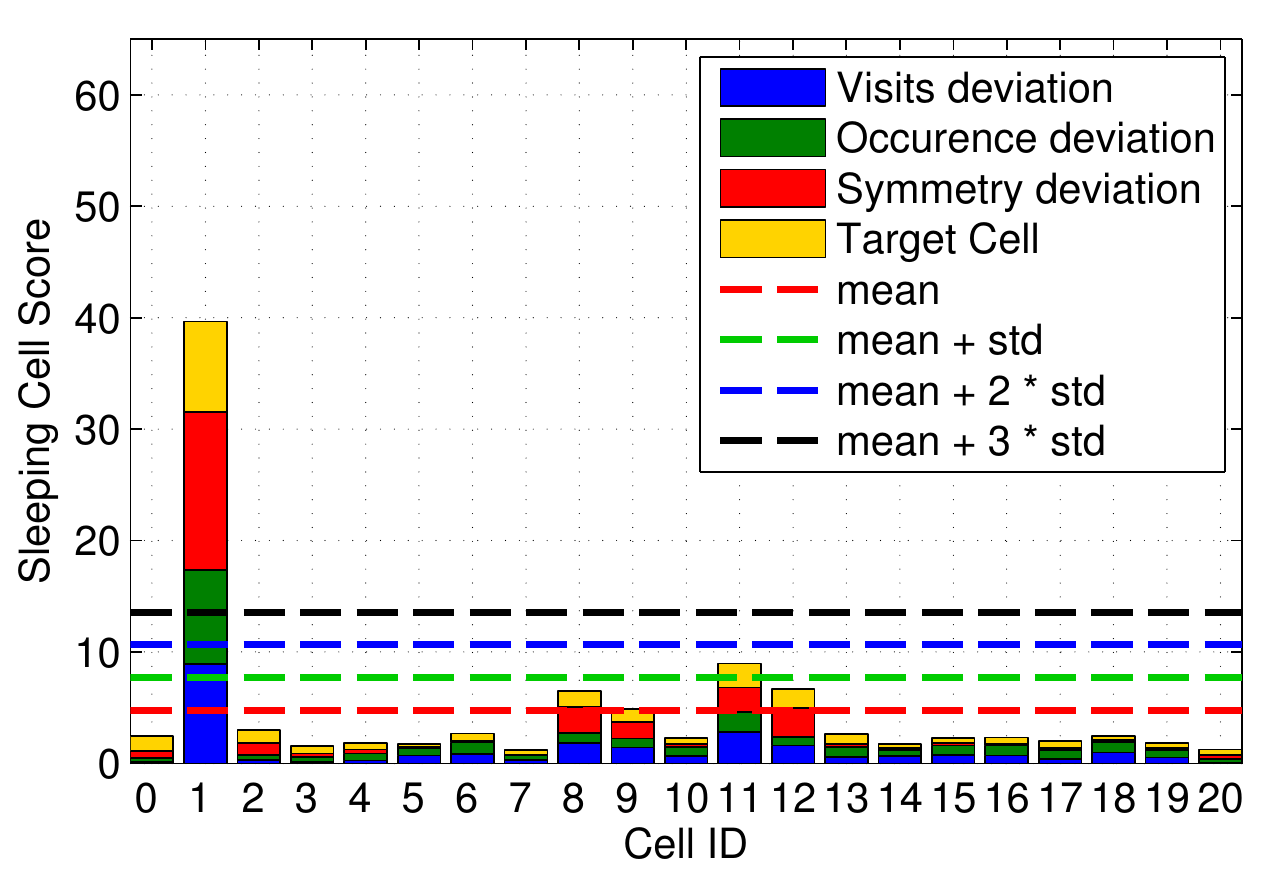}
}
\subfloat[Reference dataset sleeping cell detection histogram]{\label{fig:combined_hist_ref}\includegraphics[scale=0.44]{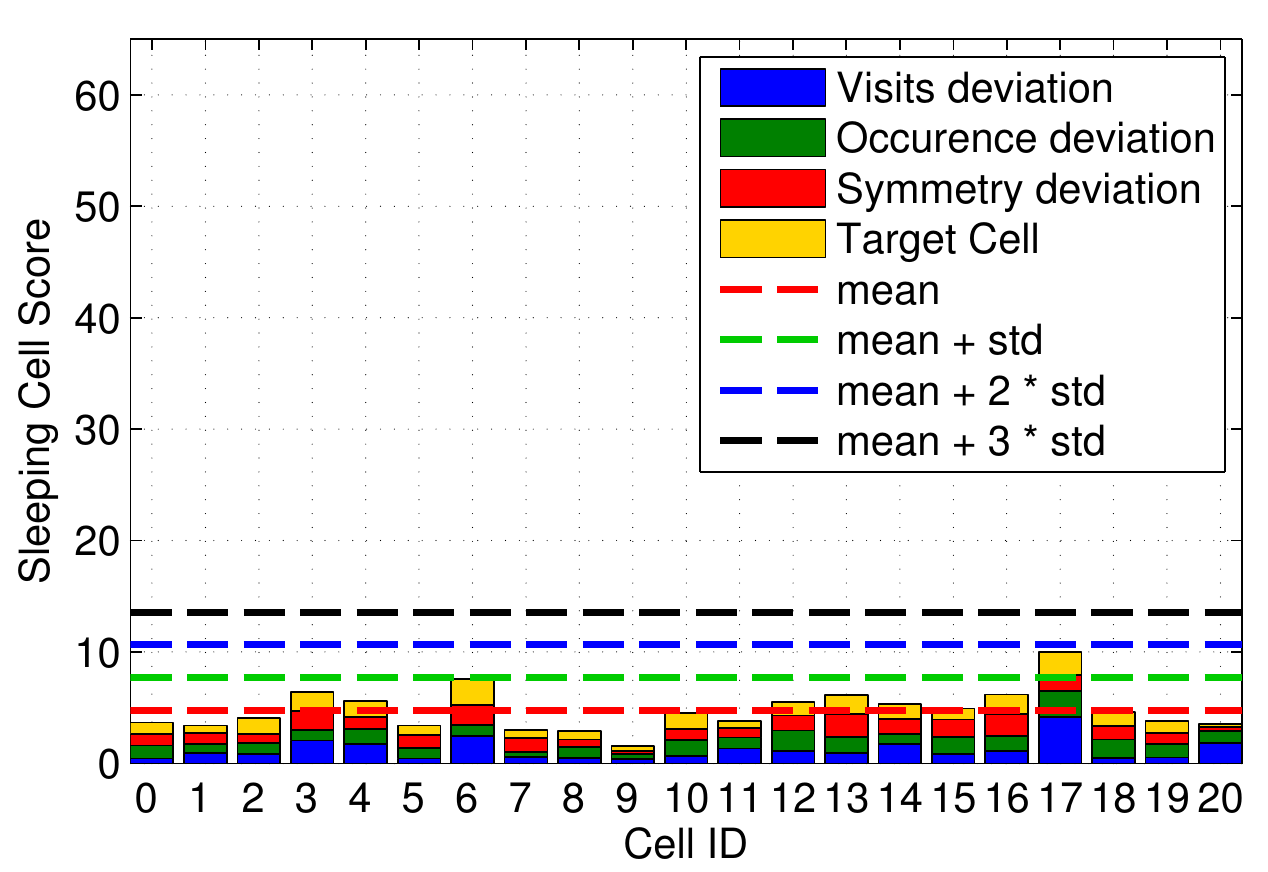}}
\\
\subfloat[Problematic dataset heat map]{\label{fig:combined_heat_map_prob}\includegraphics[scale=0.42]{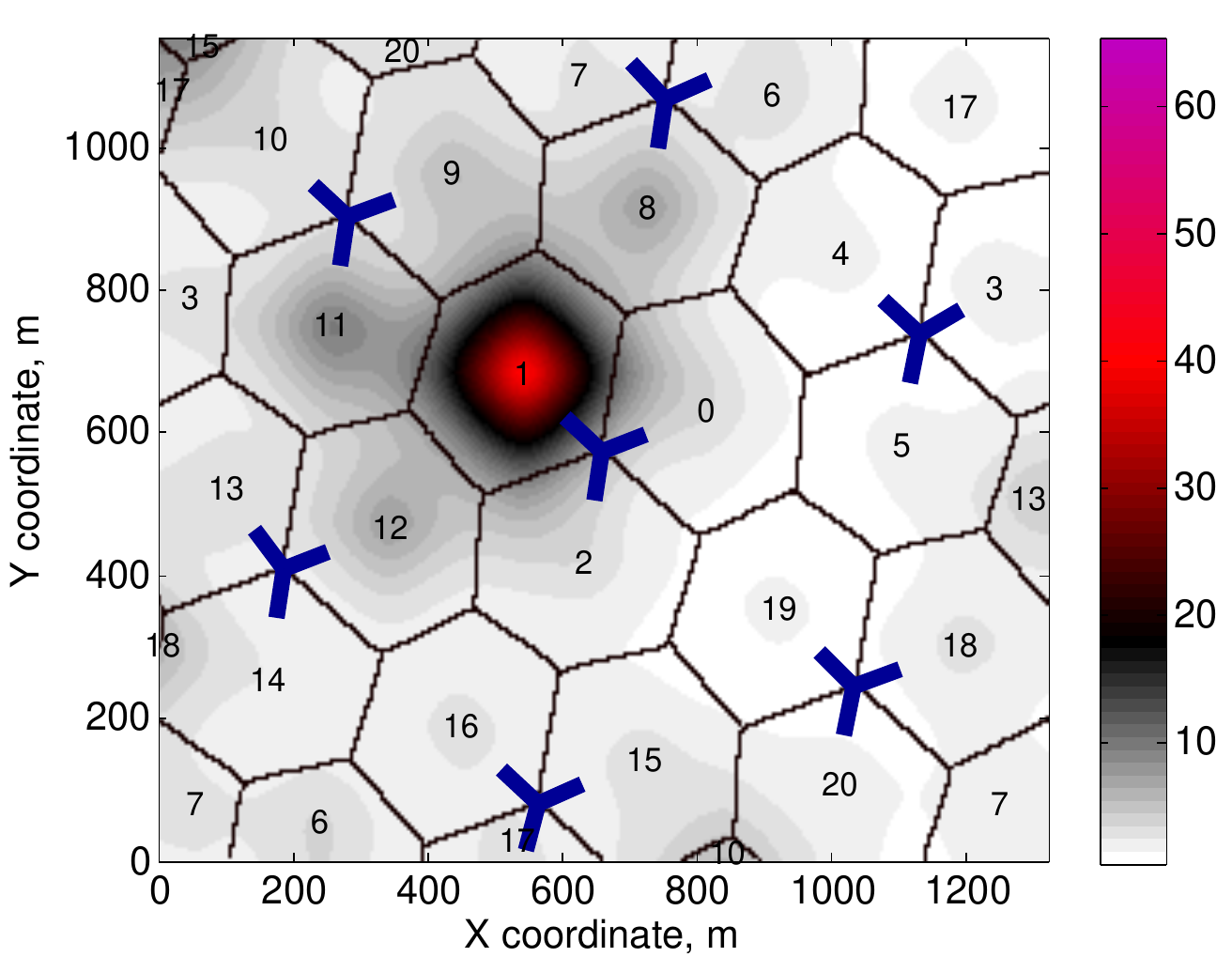}
}
\subfloat[Reference dataset heat map]{\label{fig:combined_heat_map_ref}\includegraphics[scale=0.42]{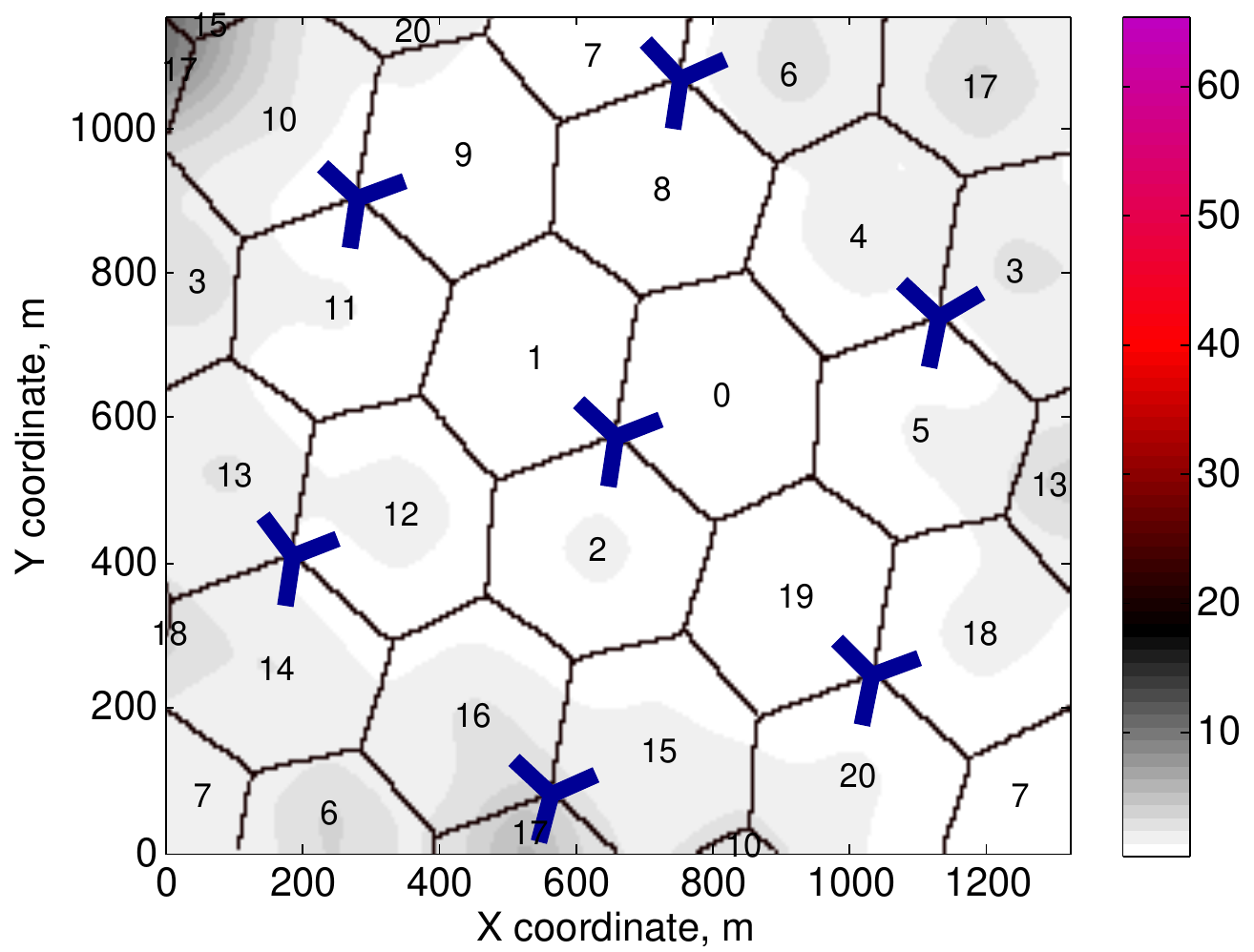}
}
\caption{Results of sleeping cell detection for amplified combined method}
\label{fig:res_combined_prob}
\end{figure}

The idea of this method is to create a cumulative sleeping cell detection histogram based on the results from all 4 post-processing methods described above. The resulting amplified \ac{sc} histogram is  shown in Fig. \ref{fig:res_combined_prob}. Cell 1 has sleeping cell score well over $\mu + 3*\sigma$ threshold. Neighboring cells 8, 9, 11, 
12 also have increased sleeping cell scores comparing to other cells.
Reference data used as testing also demonstrates stability of the 
combined approach -- no false alarms are triggered. Though, it can be seen that usage of target cell sub-call method introduces some noise.
It is important to note that post-processing methods are applied with equal weights. However, it is possible to emphasize more accurate method by increasing its weight, and penalize the unreliable, by reducing its weight. Though, selection of optimal weights is a matter of a separate study and is not discussed in this article.

\subsection{Comparison of Algorithms and Performance Evaluation}
\label{subsec:res_perf_eval}
\begin{figure}
\centering
\subfloat[Performance Measures of Algorithms]{\label{fig:perf_measures}\includegraphics[scale=0.7]{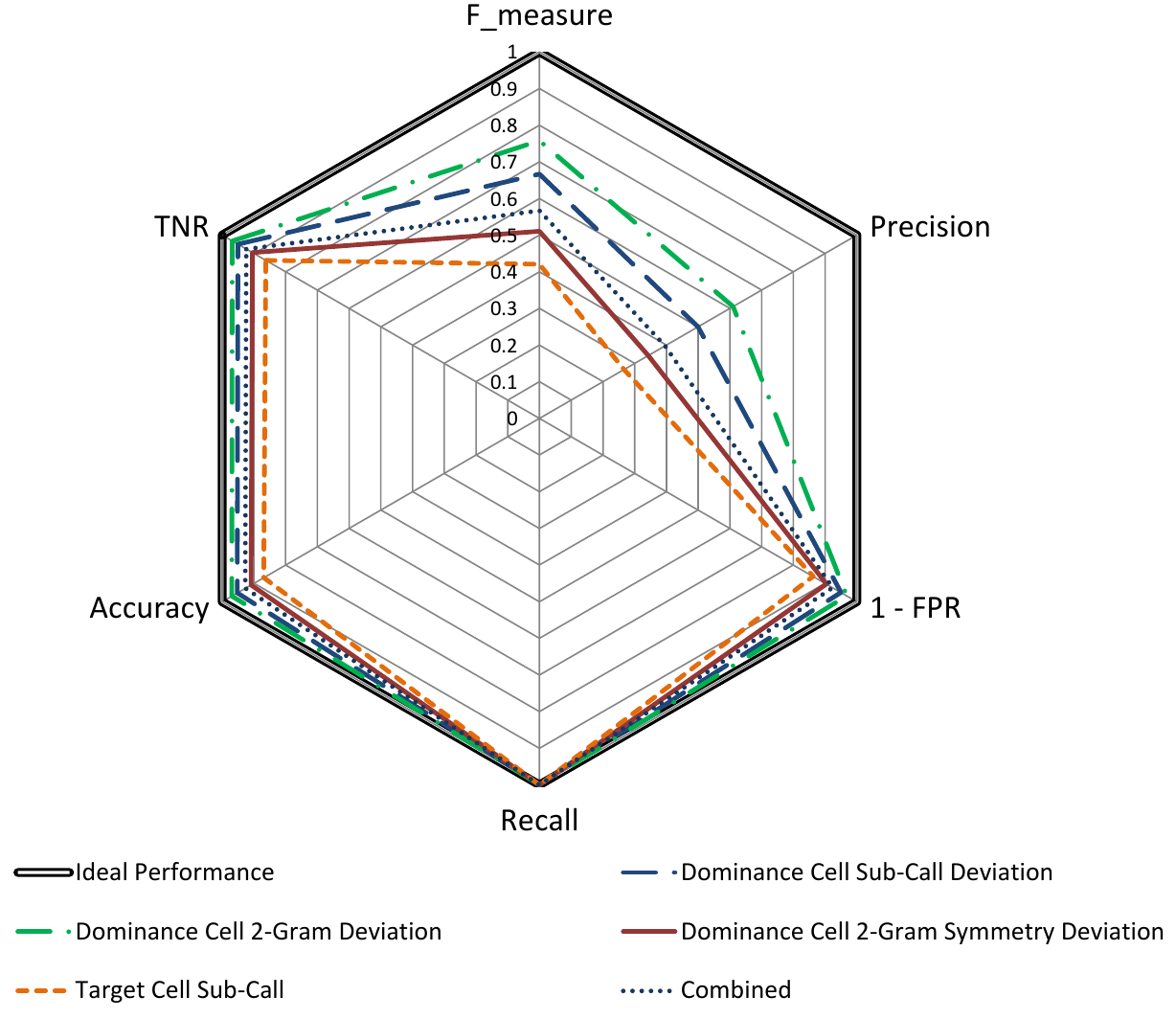}
}
\\
\vspace{10pt}%
\subfloat[ROC curve of sleeping cell detection framework]{\label{fig:roc_auc}\includegraphics[scale=0.44]{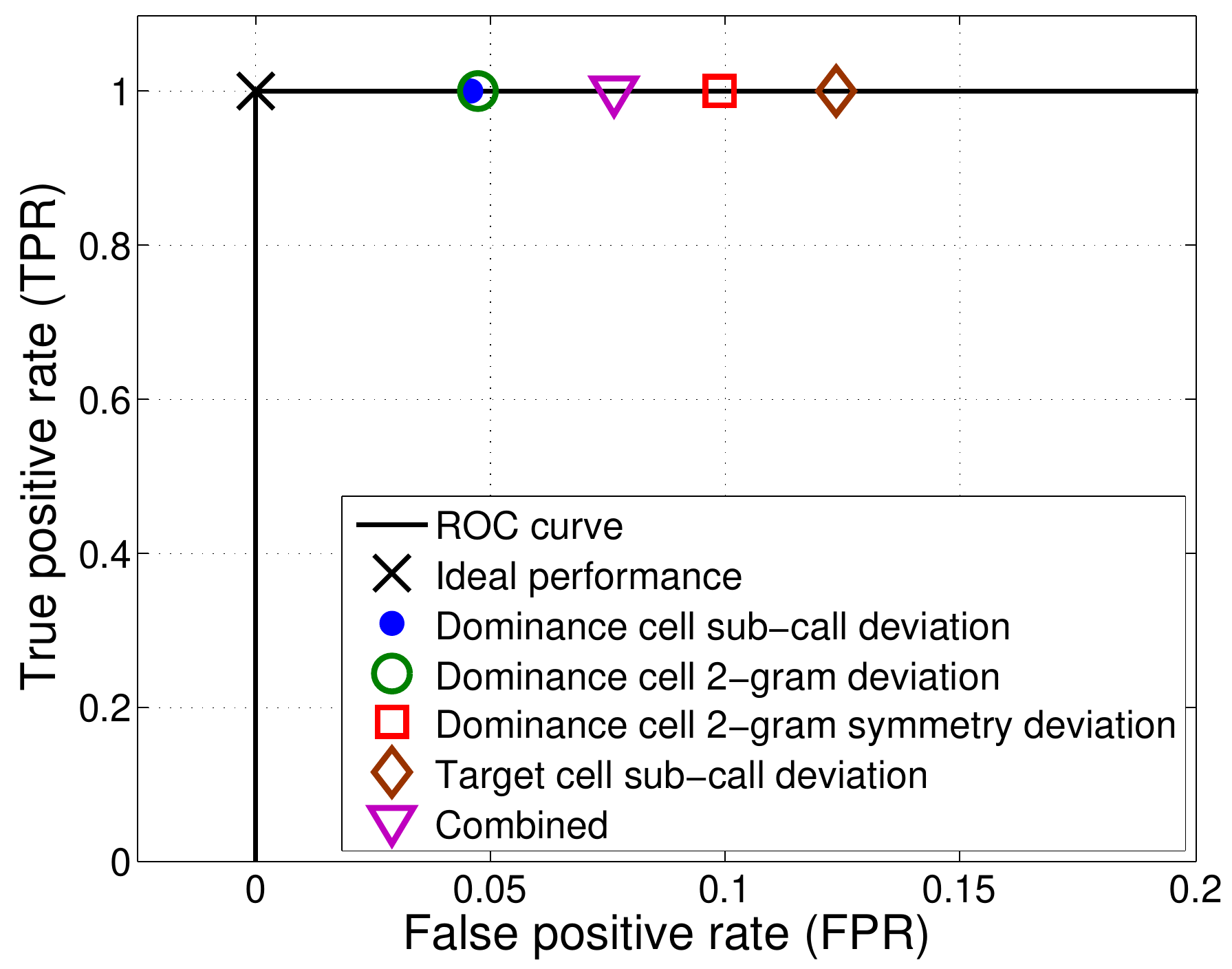}
}
\caption{Performance measures for comparison of sleeping cell detection algorithms}
\label{fig:alg_compare}
\end{figure}

The post-processing methods discussed above have their own advantages 
and disadvantages. Traditional data mining metrics, discussed in Section \ref{subsub:perf_meas}, are applied for quantitative comparison of sleeping cell detection methods, Fig. \ref{fig:perf_measures}. Ideal performance is presented with the solid double black line, and corresponds to the maximum area of the hexagon. Formally, according to the values of the metrics, Dominance Cell 2-gram Deviation and Dominance Cell Sub-call Deviation methods, demonstrate better 
performance than other post-processing techniques. However, high false positive rate for Dominance Cell 2-gram Symmetry Deviation and Target Cell Sub-call methods does not necessarily mean that these methods are worse. The reason is that neighboring cells of cell 1 exceed the 3$\sigma$ threshold. This happens because adjacent cells are not completely independent, and are affected by malfunction in one of the neighbors. 
Thus, Dominance Cell 2-gram Symmetry Deviation and Target Cell Sub-call methods can be treated as more sensitive than the others. The observed behavior emphasizes that amplification should be complemented by some other ways to to take network topology into account. However this is a subject for further study.
\ac{roc} curve of of the designed sleeping cell detection algorithm is 
presented in Fig. \ref{fig:roc_auc}. The proposed framework is able to create such a projection of the \ac{mdt} data, that in the new space 
normal data and anomalous data points are fully separable and do not overlap. Hence, the suggested data mining framework for sleeping cell detection is successful, and for reduction of false alarm rate it is necessary to invent a better separation rule, than 3$\sigma$ deviation from mean \ac{sc} score. 

\begin{figure}
\centering
\subfloat[Distances in original - non amplified approach]{\label{fig:heur_dist_orig}\includegraphics[scale=0.7]{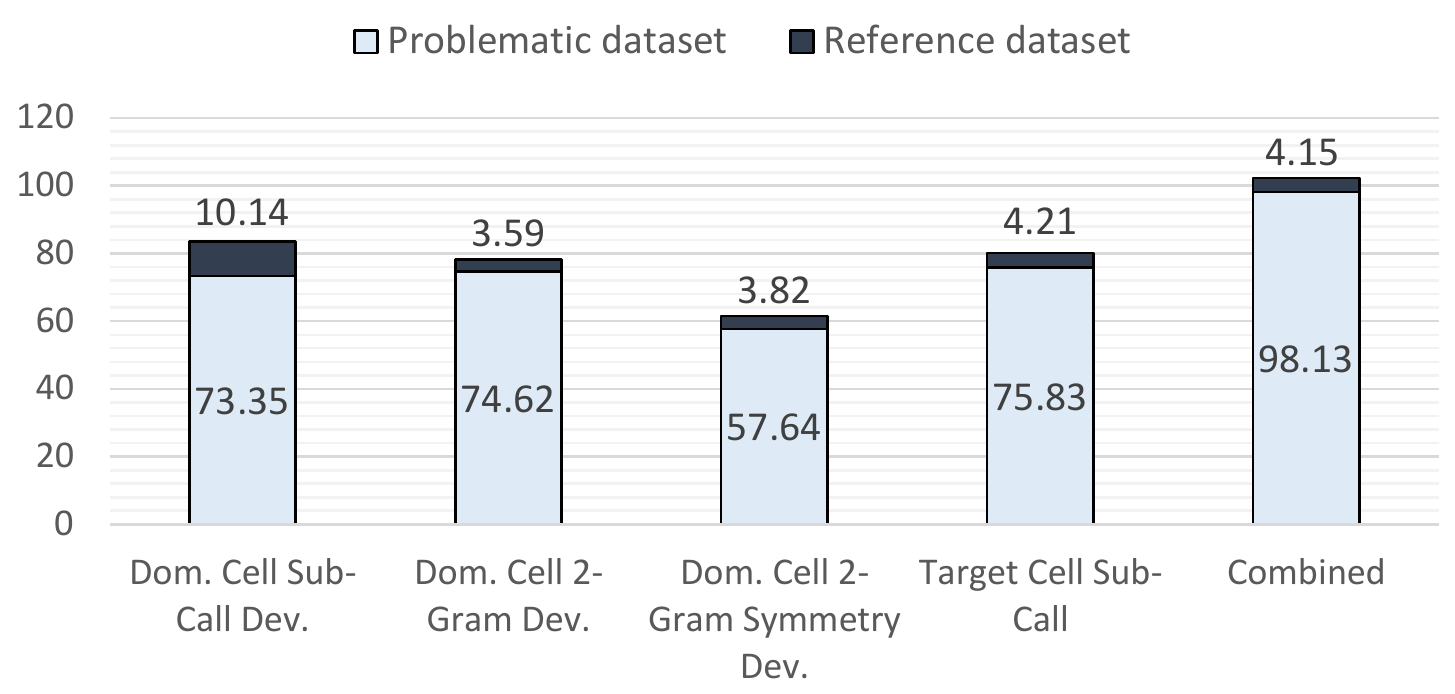}
}\\
\subfloat[Distances in amplified approach]
{\label{fig:heur_dist_amplif}\includegraphics[scale=0.7]{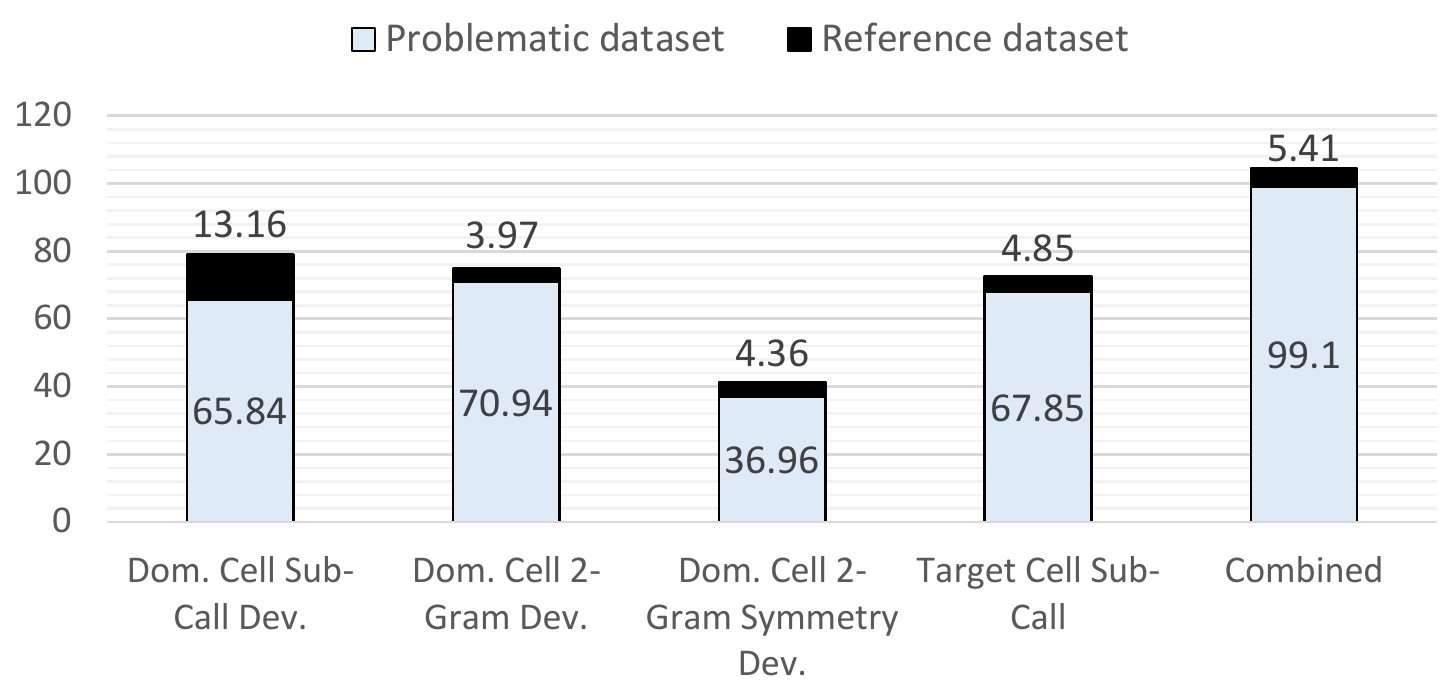}
}
\caption{Heuristic performance comparison of algorithms}
\label{fig:alg_compare_heur}
\end{figure}

Another method for comparison of post-processing algorithms is a heuristic approach described in Section \ref{subsub:perf_meas}. According to this method, more accurate post-processing algorithm is the one, which has the smallest distance to the ideal solution point for either problematic or error-free case. Cumulative distances for different algorithms in non-amplified and amplified cases are presented in Fig. \ref{fig:heur_dist_orig} and Fig. \ref{fig:heur_dist_amplif} correspondingly.
It can be seen that Dominance Cell 2-Gram Symmetry Deviation method 
has the smallest distance from the ideal detection case. Thus, from 
perspective of the heuristic performance evaluation approach this method outperforms other post-processing methods.

\section{Conclusions}
\label{sec:conclusions}
This article presents a novel sleeping cell detection framework based on knowledge mining paradigm. \ac{mdt} reports are used for the detection of a random access channel malfunction in one of the network cells. Experimental setup implements a simulated \ac{lte} network, used to generate a diverse statistics base with several thousands of user calls and tens of thousands of \ac{mdt} samples. Investigated type of sleeping cell problem is rather complex, and detection of this problem has never been studied before. 

The designed knowledge mining framework is semi-supervised and has centralized architecture from perspective of self-organizing networks. The heart of the developed detection framework is the analysis of sequences with N-gram method in the series of user event-triggered measurement \ac{mdt} reports. Data pre-processing with sliding window transformation method allows to make the statistics base more reliable through standardization of the input event sequences. 2-gram analysis is used to convert sequential data to numeric format in the new feature space. To simplify analysis of the data in the new space, dimensionality reduction with minor component analysis method is applied. \ac{knn} anomaly score detection algorithm is used to find the outliers in the data and using this information, anomalous data points are converted with post-processing to the knowledge about location of the problematic regions in the network. Comparison of different location mapping post-processing methods is done, additionally, so called amplification is used to take into account neighbor relations between cells and network topology, for improvement of sleeping cell detection performance. 

Results demonstrate, that the suggested framework allows for efficient detection of the random access sleeping cell problem in the network. Evaluation shows that post-processing method named Dominance Cell 2-Gram Symmetry Deviation demonstrates the best combination of performance results. Amplification also proves to be the very efficient approach for improvement of the detection quality.
Results of this work lay grounds and suggest exact methods for 
building advanced performance monitoring systems in modern mobile 
networks. One of the possible directions in this area is extensive usage 
of data mining techniques in general, and anomaly detection in 
particular. New systems of network maintenance would allow to address 
growing complexity and heterogeneity of modern mobile networks, and 
especially \ac{5g}.

Future work in this field includes validation of the developed system 
in more complex scenarios, detection of several or different types of 
malfunctions, and substitution of semi-supervised approach with 
unsupervised. The ultimate goal is to achieve accurate and timely detection of different sleeping cell types in highly dynamic mobile network environments. Obviously, low level of false alarms must be 
supported, and at the same time significant increase of computational 
complexity should be avoided.

\section*{Acknowledgments}
Authors would like to thank colleagues from Magister Solutions, Nokia 
and University of Jyv\"{a}skyl\"{a} for collaboration, their valuable 
feedback regarding this research, and peer reviews. Work on this study 
has been partly funded by MIPCOM project, Graduate School in 
Electronics, Telecommunications and Automation (GETA), and Doctoral 
Program in Computing and Mathematical Sciences (COMAS).
\bibliographystyle{spbasic}
\bibliography{bibliography_all}
\newpage

\end{document}